\documentclass[epj,nopacs,fleqn]{svjour}
\pdfoutput=1
\usepackage{slashed,graphicx,amsmath,amssymb,cite,url,hyperref,dsfont,%
  subfigure,graphicx,placeins}
\usepackage[utf8]{inputenc}
\hypersetup{bookmarks=true,unicode=true,pdftoolbar=true,pdfmenubar=true,
  pdffitwindow=false,pdfstartview={FitH},
  pdfnewwindow=true,colorlinks=true,
  linkcolor=blue,citecolor=magenta,filecolor=magenta,urlcolor=cyan}

\newcommand{\beq}{\begin{equation}}
\newcommand{\eeq}{\end{equation}}
\newcommand{\bea}{\begin{eqnarray}}
\newcommand{\eea}{\end{eqnarray}}
\newcommand{\be}{\begin{equation}}
\newcommand{\ee}{\end{equation}}
\def\bsp#1\esp{\begin{split}#1\end{split}}
\def\bpm{\begin{pmatrix}}
\def\epm{\end{pmatrix}}

\newcommand{\lb}{\left}
\newcommand{\rb}{\right}

\begin{document}

\title{Realistic simplified gaugino-higgsino models in the MSSM}

\author{
  Benjamin Fuks\inst{1,2,3}\thanks{\color{blue}fuks@lpthe.jussieu.fr},
  Michael Klasen\inst{4}\thanks{\color{blue}michael.klasen@uni-muenster.de},
  Saskia Schmiemann\inst{4}\thanks{\color{blue}saskia.schmiemann@uni-muenster.de} and
  Marthijn Sunder\inst{4}\thanks{\color{blue}mpasunder@uni-muenster.de}
}

\institute{
  Sorbonne Universit\'es, Universit\'e Pierre et Marie Curie
  (Paris 06), UMR 7589, LPTHE, F-75005 Paris, France
  \and
  CNRS, UMR 7589, LPTHE, F-75005, Paris, France
  \and
  Institut Universitaire de France, 103 boulevard Saint-Michel,
  F-75005 Paris, France
  \and
  Institut f\"ur Theoretische Physik,
  Westf\"alische Wilhelms-Universit\"at
  M\"unster,\\ Wilhelm-Klemm-Stra\ss{}e 9, D-48149 M\"unster, Germany
}

\date{}

\abstract{We present simplified MSSM models for light neutralinos and charginos with realistic
 mass spectra and realistic gaugino-higgsino mixing, that can be used in experimental searches
 at the LHC. The formerly used naive approach of defining mass spectra and mixing matrix elements
 manually and independently of each other does not yield genuine MSSM benchmarks. We suggest the
 use
 of less simplified, but realistic MSSM models, whose mass spectra and mixing matrix elements are
 the result of a proper matrix diagonalisation. We propose a novel strategy targeting the
 design of such benchmark scenarios, accounting for user-defined constraints in terms of masses and particle
 mixing. We apply it to the higgsino case and implement a scan in the four relevant
 underlying parameters $\{\mu, \tan \beta, M_{1}, M_{2}\}$ for a given set of light neutralino
 and chargino masses. We define a measure for the quality of the obtained benchmarks,
 that also includes criteria to assess the higgsino content of the resulting charginos and
 neutralinos. We finally discuss the distribution of the
 resulting models in the MSSM parameter space as well as their implications for supersymmetric
 dark matter phenomenology.}

\titlerunning{Realistic simplified gaugino-higgsino models in the MSSM}
\authorrunning{B.~Fuks {\it et al.}}


\maketitle
\flushbottom
\vspace*{-12.5cm}
\noindent \today\\
\small{MS-TP-17-14}
\vspace*{10.7cm}

\section{Introduction} \label{sec:1}

Supersymmetry (SUSY) is one of the most popular theories beyond the Standard Model (SM) of
particle physics. Extending the Poincar\'e algebra by relating the fermionic and
bosonic degrees of freedom of the theory, supersymmetry provides a solution to
many of the shortcomings and limitations of the Standard Model. In particular,
supersymmetric theories solve the infamous hierarchy problem plaguing the
Standard Model, feature gauge coupling unification at high energy and generally
include a natural explanation for the presence of dark matter in the universe.
Consequently, supersymmetry searches constitute a significant part of the LHC physics program.

Up to now, no evidence for supersymmetry has been found. Limits on the
masses of the supersymmetric partners of the Standard Model particles are
consequently pushed to higher and higher energy scales. Most of these results
have, however, been derived either in the framework of the minimal
supersymmetric realisation, known as the Minimal Supersymmetric Standard Model
(MSSM)~\cite{Nilles:1983ge,Haber:1984rc}, or within MSSM-inspired simplified
models for new physics~\cite{Alwall:2008va,Alwall:2008ag,Alves:2011wf,Calibbi:2014lga}.

Simplified models are effective Lagrangian descriptions minimally extending the
Standard Model in terms of new particles and interactions. They have been
designed as useful tools for the characterisation of new phenomena, allowing for
the reinterpretation of the results in a straightforward manner thanks to a
reduced set of degrees of freedom. In the context of MSSM-inspired simplified models, the
experimental attention was initially mainly focused on the analysis of
signatures that could originate from the strong production of squarks and
gluinos, the corresponding cross sections being expected to be larger by virtue
of the properties of the strong interaction. LHC null results have implied that
severe constraints are now imposed on the masses of these strongly interacting
superpartners. In particular, the analysis of about 36~fb$^{-1}$ of LHC
collision data at a centre-of-mass energy of 13~TeV pushes the lower bounds on
these masses far into the multi-TeV regime~\cite{Aaboud:2017ayj,%
Aaboud:2017wqg,Aaboud:2017bac,Aaboud:2017nfd,Aaboud:2017dmy,Aaboud:2017ejf,%
Aaboud:2017hdf,Sirunyan:2017cwe,Sirunyan:2017fsj,%
Sirunyan:2017kqq,Sirunyan:2017mrs,Sirunyan:2017uyt,Sirunyan:2017yse,%
Sirunyan:2017wif,Sirunyan:2017kiw,Sirunyan:2017xse,Sirunyan:2017qaj}.
Processes involving the production of a pair of electroweak superpartners
(neutralinos, charginos and sleptons) have also been considered for some
time. The electroweak nature of these processes yields, however,
smaller production rates and subsequently softer bounds on the corresponding
masses~\cite{Aaboud:2017nhr,Sirunyan:2017qaj,Sirunyan:2017lae,Sirunyan:2017obz}.
Neutralinos, charginos and sleptons of a few hundreds of GeV are indeed still
allowed by current data.

We focus in this work on simplified models describing electroweak gauginos
and higgsinos  and their dynamics. Recent searches of both ATLAS and CMS are in general
interpreted within the framework of two sets of simplified models. In the first
case, the Standard Model is extended by a set of mass-degenerate pure wino
states, and the lightest superpartner is a pure bino state. The winos are then
assumed to decay either into a system made of a bino and a weak gauge or Higgs boson,
regardless of the fact that these decays are strictly speaking not allowed by
supersymmetric gauge invariance, or into a bino and jets or leptons via
intermediate off-shell sfermions. When the gaugino-higgsino mixing is not negligible
and the mass splitting between the lightest states is sufficiently large, the
decays to weak gauge and Higgs bosons become
allowed and provide opportunities to obtain bounds on the MSSM parameter space.
The strength of the constraints then depends on the mixing and the
mass splitting~\cite{Bharucha:2013epa}.
On the other hand, heavier higgsino-like electroweakinos decay dominantly into
lightest neutralinos and weak gauge bosons, thanks to their
mixing with the gauginos, but only if the channels are kinematically
accessible. In compressed mass scenarios, the corresponding
experimental  searches
rely on the detection of the soft decay products of the gauge bosons, {\it e.g.}
low transverse-momentum opposite-charge leptons of the same flavour (electrons or
muons) \cite{Aaboud:2017leg,Sirunyan:2018iwl}.
The second set of models under consideration is
inspired by gauge-mediated supersymmetry breaking scenarios~\cite{Dine:1981gu,%
Nappi:1982hm,AlvarezGaume:1981wy,Dine:1993yw,Dine:1994vc,Dine:1995ag,%
Giudice:1998bp}, in which the lightest superpartner is the gravitino. This
simplified model additionally contains two neutral and one charged higgsino state,
which are quasi mass-degenerate. They hence decay into a gravitino and a neutral gauge
or Higgs boson, together with possibly accompanying undetected soft objects.

In all of the above approaches to MSSM-inspired simplified models for
the gaugino-higgsino sector, one naively ignores all interrelationships between
the masses of the neutralinos and the charginos and the features of the
associated mixing matrices through their respective dependence on the free parameters
in the MSSM Lagrangian. Starting from the MSSM, the neutralinos and
charginos that are not of interest are decoupled by imposing the
corresponding mixing matrix elements to be vanishing and their masses to be
very large. On the other hand, the masses of the relevant neutralinos and charginos are
fixed by hand to the desired values, independently of the corresponding elements in the mixing
matrices that are set to 0, 1, or $\pm 1/\sqrt{2}$ (in the higgsino case). This
approach is justified by the assumption that the MSSM has sufficiently many free parameters
to reproduce such a pattern closely enough, which is particularly true when one
considers the extra freedoms originating from the loop corrections.

In certain configurations, {\it e.g.} when the lightest states
are nearly degenerate pure higgsinos and the gauginos are decoupled,
this simple method works quite well. However, when one targets next-to-minimal
simplified models where a mass splitting between the second-lightest state and its
neighbours is introduced, some amount of mixing between the different gaugino and
higgsino fields must be included in order to maintain viability with
respect to the initial MSSM motivation. This concerns in particular identities
guaranteed by gauge invariance and/or supersymmetry that could be violated when
one tweaks by hand masses and mixing matrix elements, like in the
above-mentioned wino set of simplified models.
Such non-minimal setups are already probed by both LHC collaborations in their
searches for supersymmetry~\cite{Aaboud:2017leg,Sirunyan:2018iwl}. It is
therefore important to interpret the
results in meaningful benchmark scenarios where supersymmetry and gauge
symmetries are preserved, allowing in this way only for theoretically-relevant
interpretations.

In this work, we present simplified MSSM models for light neutralinos and charginos with
realistic mass spectra and realistic gaugino-higgsino mixing, that can be used, {\it e.g.}, in
experimental searches at the LHC. Starting from the MSSM without additional $CP$-violation, we
design our simplified model by decoupling all coloured superpartners as well as the sleptons and
the sneutrinos. The gaugino-higgsino sector is thus described, at tree-level, by four parameters
that are the bino and wino mass parameters $M_1$ and $M_2$, the supersymmetric higgs(ino)
off-diagonal mass parameter $\mu$, and the ratio of the vacuum expectation values of the neutral
components of the two Higgs doublets $\tan\beta$. We then define a strategy to efficiently scan
this four-dimenensional parameter space for given sets of light neutralino and chargino masses,
that also allows to maximise the gaugino or higgsino content, couplings to certain sparticles etc.
This procedure therefore allows to find approximate solutions for simplified MSSM models that
have a realistic and properly defined gaugino-higgsino sector in contrast to many of the
overly simplified models studied so far.

The remainder of this paper is organised as follows. We first review in Sec.\ \ref{sec:2} the
MSSM chargino-neutralino sector, discuss its analytic symmetries, and study the spectra and
decompositions of the physical states after numerical diagonalisation of the neutralino and
chargino mass matrices. In Sec.~\ref{sec:3}, we describe our strategy to scan the four-dimensional
MSSM parameter space, define a quality measure for the goodness of our fit to the desired
simplified
model, and indicate how our scan strategy can be generalised. In Sec.~\ref{sec:4}, we present
a case study for higgsino-like light neutralinos and charginos, analyse their representation in
the MSSM parameter space, and investigate the implications for the Higgs-stop sector as well
as the phenomenology of supersymmetric dark matter. Our conclusions are given in Sec.\
\ref{sec:5}.

\section{Theoretical definitions} \label{sec:2}

The simplified model that we investigate in this work takes the
gaugino-higgsino sector from the MSSM in all its complexity, as it is defined by
supersymmetry and gauge invariance. In other words, we compute all elements of
the neutralino and chargino mixing matrices and the physical mass spectrum
through a proper diagonalisation of the relevant mass matrices at tree level. In
our procedure, the mass spectrum of the neutralinos and charginos is thus not
treated independently from their couplings, as it has been done previously in
(overly) simplified models. By decoupling other supersymmetric particles,
the model does, however, still not become overly complex, and this partly
justifies that we neglect higher-order effects. The latter are nevertheless not
so relevant for our purpose, the idea being to design models
closely enough reproducible in the MSSM.

\subsection{MSSM chargino-neutralino sector} \label{ssec:2.1}

In the MSSM and at tree-level, the gaugino-higgsino (or equivalently
neutralino-chargino) sector is defined by four
parameters
\be
  \{\mu,\tan \beta, M_{1}, M_{2}\} \ ,
\label{eq:prms}\ee
that are the off-diagonal Higgs(ino) mass parameter, the ratio of the vacuum
expectation values of the neutral components of the two doublets of Higgs fields
and the two soft supersymmetry-breaking electroweak gaugino mass parameters,
respectively.

The $\mu$ parameter originates from the MSSM superpotential
($\mathcal{W}_{\text{MSSM}}$). It reads, when we assume that the superpotential
contains only $R$-parity conserving terms,
\be\bsp
 \mathcal{W}_{\text{MSSM}} = &\
     \mu H_{1} \cdot H_{2}
   - y^{e}_{ij} H_{1}\cdot L_{i}E_{j}
   - y^{d}_{ij} H_{1}\cdot Q_{i}D_{j} \\ &\quad
   - y^{u}_{ij} Q_{i} \cdot H_{2}U_{j} \ ,
\esp\label{eq:ino-w}\ee
where $H_1$ and $H_2$ denote the two weak doublets of Higgs superfields.
$Q$, $L$, and $U$, $D$ and $E$ are the two weak doublets and three weak singlets
of quark and lepton superfields, respectively. Expanding the
superpotential
$\mathcal{W}_{\text{MSSM}}$ in terms of the component fields of the various
superfields, it includes in particular an off-diagonal mass term proportional to
$\mu$ for the two higgsino fields $\widetilde H_1$ and $\widetilde H_2$.
The second parameter in Eq.~\eqref{eq:prms} is defined as the ratio of the
vacuum expectation values of the scalar components $h_1$ and $h_2$ of the two
Higgs superfields
\be
 \tan\beta = \frac{v_{2}}{v_{1}}
\ee
with
\be
 \lb\langle h_{1} \rb\rangle = \frac{1}{\sqrt{2}}\bpm v_{1} \\ 0\epm
 \qquad{\rm and}\qquad
 \lb\langle h_{2} \rb\rangle = \frac{1}{\sqrt{2}}\bpm 0 \\ v_{2} \epm \ ,
\ee
the non-vanishing values of $v_1$ and $v_2$ giving rise to the spontaneous
breaking of the electroweak symmetry,
$SU(2)_{\text{L}}\times U(1)_{\text{Y}}\rightarrow U(1)_{\text{EM}}$.
Since supersymmetry has not yet been observed, it must be a broken symmetry. As
usual, we remain agnostic of which mechanism is invoked to break supersymmetry,
and thus explicitly include in the MSSM Lagrangian soft supersymmetry-breaking
interaction terms that leave the gauge symmetries intact and that do not
introduce any new quadratic divergences at the loop-level. Among the allowed
supersymmetry breaking terms, the bino ($\widetilde{B}$) and wino
($\widetilde{W}$) mass terms are the only ones relevant for our work,
\be
 \label{eq:ino-soft} 
 \mathcal{L}^{\text{MSSM}}_{\text{soft}} =
 - \frac{1}{2}\lb( M_{1} \widetilde{B} \widetilde{B}
    + M_{2} \widetilde{W}_{i} \widetilde{W}^{i} + \text{h.c.}\rb) + \ldots \ .
\ee

The chargino mass eigenvalues are obtained by diagonalising the chargino mass
matrix ${\bf X}$ that can be extracted from Eq.~\eqref{eq:ino-w} and
Eq.~\eqref{eq:ino-soft}. This matrix is given, in the $(i \widetilde W^-,
\widetilde H^-_1)$ and $(i \widetilde W^+, \widetilde H^+_2)$ basis, by
\be\label{eq:chargino-1}
 \mathbf{X} =
   \bpm M_{2}&\sqrt{2} M_Ws_\beta \\ \sqrt{2} M_Wc_\beta & \mu \epm \ ,
\ee
where $M_W$ stands for the mass of the $W$-boson and where we have introduced
the $c_\beta$ and $s_\beta$ notations for the cosine and sine of the $\beta$
angle, respectively. This matrix can be diagonalised by means of two unitary
rotation matrices ${\cal U}$ and ${\cal V}$,
\be\label{eq:chargino-2} 
 {\rm diag} (M_{\tilde{\chi}^\pm_1}, M_{\tilde{\chi}^\pm_2})  =
     \mathcal{U}^{*} \mathbf{X} \mathcal{V}^{-1} \ ,
\ee
where $M_{\tilde{\chi}^\pm_1} < M_{\tilde{\chi}^\pm_2}$ are the masses of
the two chargino states. The ${\cal U}$ and ${\cal V}$ mixing matrices
respectively relate the negatively-charged and positively-charged
gaugino-higgsino basis to the physical chargino mass basis $(\chi^\pm_1,
\chi^\pm_2)$,
\be
  \bpm \chi^-_1 \\ \chi^-_2 \epm =
    {\cal U} \bpm i \widetilde W^- \\ \widetilde H^-_d \epm
  \quad \text{and}\quad
  \bpm \chi^+_1 \\ \chi^+_2 \epm =
    {\cal V} \bpm i \widetilde W^+ \\ \widetilde H^+_u \epm \ .
\ee

Similarly, in the neutral sector the neutralino mass matrix can
be computed from Eq.~\eqref{eq:ino-w} and Eq.~\eqref{eq:ino-soft}. This matrix
can be written, in the $(i\widetilde B, i \widetilde W^3, \widetilde H_1^0,
\widetilde H_2^0)$ basis, as
\be \label{eq:neutralino-1}
 \mathbf{Y} = \bpm
    M_1               & 0            & -M_W t_W c_\beta & M_W t_W s_\beta \\
    0                 & M_2          &  M_W c_\beta     & -M_W s_\beta\\
    - M_W t_W c_\beta & M_W  c_\beta & 0                & -\mu\\
      M_W t_W s_\beta & -M_W s_\beta & -\mu             & 0
  \epm \ ,
\ee
where $t_W \equiv \tan\theta_W$ stands for the tangent of the electroweak mixing
angle. This symmetric matrix can be diagonalised by means of a single
unitary matrix ${\cal N}$,
\be \label{eq:neutralino-2} 
  {\cal N}^* {\bf Y} {\cal N}^{-1} = {\rm diag}\,
    (M_{\tilde{\chi}^0_1}, M_{\tilde{\chi}^0_2},
     M_{\tilde{\chi}^0_3}, M_{\tilde{\chi}^0_4}) \ ,
\ee
where $M_{\tilde{\chi}^0_1} < M_{\tilde{\chi}^0_2} < M_{\tilde{\chi}^0_3} <
M_{\tilde{\chi}^0_4}$ stand for the masses of the four neutralino states
$\chi_i^0$ with $i=1$, 2, 3 and 4. The mixing matrix ${\cal N}$ allows one to
relate the four physical neutralino mass eigenstates to the neutral higgsino and
gaugino interaction eigenstates,
\be
  \bpm \chi^0_1 \\ \chi^0_2 \\ \chi^0_3 \\ \chi^0_4 \epm = {\cal N}
  \bpm i \widetilde B \\ i \widetilde W^3 \\ \widetilde H_d^0 \\ 
     \widetilde H_u^0 \epm \ .
\ee
Non-trivial analytic inversions of the gaugino mass matrices have been
proposed in the past. Besides the knowledge of three gaugino masses, typically
those of one or two charginos and of two or one heavier neutralinos, they require the
choice of a value for $\tan\beta$ as well as additional information and/or
numerical consistency checks to resolve sign ambiguities~\cite{Kneur:1998gy}.

\subsection{Symmetry transformations} \label{ssec:2.2}

For a better understanding of the structure of the parameter space of our
simplified model, we discuss in this subsection two linear transformations of
the mixing matrices that affect the electroweakino couplings, but leave their
mass spectrum unchanged. These symmetries hence allow us to deduce multiple
benchmark scenarios fitting equally well a preselected mass configuration and
chargino and neutralino decomposition in terms of gaugino and higgsino
eigenstates.

We restrict our study to the case where the $\mu$, $M_{1}$ and $M_{2}$
parameters are real in order not to introduce additional sources of
$CP$-violation in the theory. However, we keep the sign of these three mass
parameters free, so that they can therefore be either positive or negative. The
mass eigenvalues of the chargino mass matrix ${\bf X}$ only depend on the
relative sign between the $\mu$ and $M_{2}$ parameters. This means that the
simultaneous flip of the signs of the $M_2$ and $\mu$ parameters,
\be \label{eq:chargsigntr}
 M_2 \to M_{2}^{\prime} = -M_{2} \qquad\text{and}\qquad
 \mu \to \mu^{\prime} = -\mu \ ,
\ee
leaves both chargino masses invariant. The chargino mixing matrices are, however,
impacted and transform as
\be \label{eq:chargsignmm} 
 {\cal U} \to \mathcal{U}^{\prime} = -\mathcal{U} \sigma_{3}
  \qquad\text{and}\qquad
 {\cal V} \to \mathcal{V}^{\prime} = \mathcal{V} \sigma_{3} \ ,
\ee
where $\sigma_3$ is the third Pauli matrix. In general, these two sign flips
also lead to effects on the neutralino mass spectrum, unless one extends the
transformation of Eq.~\eqref{eq:chargsigntr} as
\be \label{eq:neutsigntr}\bsp
  & M_1 \to M_{1}^{\prime} = -M_{1} \ , \\
 & M_2 \to M_{2}^{\prime} = -M_{2} \ , \\
 & \mu \to \mu^{\prime} = -\mu \ .
\esp\ee
The neutralino masses are thus left invariant by the transformation of
Eq.~\eqref{eq:neutsigntr}, that modifies the neutralino mixing matrix ${\cal N}$
as
\be \label{eq:neutsignmm} 
 {\cal N} \to \mathcal{N}^{\prime} = i \mathcal{N} \bpm
   \phantom{-}1 & \phantom{-}0 & \phantom{-}0 & \phantom{-}0\\
   \phantom{-}0 & \phantom{-}1 & \phantom{-}0 & \phantom{-}0\\
   \phantom{-}0 & \phantom{-}0 &          - 1 & \phantom{-}0\\
   \phantom{-}0 & \phantom{-}0 & \phantom{-}0 &          - 1\\
  \epm \ .
\ee

On different grounds, the inversion of $\tan\beta$,
\be \label{eq:tanbtr} 
 \tan\beta \to [\tan \beta]^{\prime} =  \frac{1}{\tan \beta} \ ,
\ee
also leaves the chargino and neutralino mass spectrum invariant. The mixing
matrices $\mathcal{U}$ and $\mathcal{V}$ are, however, interchanged, as are the
decompositions of the Weyl fields $\chi^{+}_{i}$ and $\chi^{-}_{i}$ in terms of
their gaugino and higgsino content,
\be
 {\cal U} \to \mathcal{U}^{\prime} = \mathcal{V} \bpm 0 & 1\\ 1 & 0\epm
  \qquad\text{and}\qquad
 {\cal V} \to \mathcal{V}^{\prime} = \mathcal{U}  \bpm 0 & 1\\ 1 & 0\epm \ .
\ee
The total gaugino-higgsino content of the Dirac chargino spinors is, however,
unaffected. As mentioned above, the transformation of Eq.~\eqref{eq:tanbtr} also
leaves the neutralino mass eigenvalues invariant. The neutralino mixing matrix
${\cal N}$ is in contrast modified. The inversion of $\tan\beta$ physically
interchanges the roles of $\widetilde{H}^{0}_{1}$ and $\widetilde{H}^{0}_{2}$,
so that the decomposition of the neutralinos in terms of the two higgsino states
is swapped with an extra sign flip,
\be
 \label{eq:neuttanbmm} 
 {\cal N} \to \mathcal{N}^{\prime} =   \mathcal{N} \bpm
   \phantom{-}1 & \phantom{-}0 & \phantom{-}0 & \phantom{-}0\\
   \phantom{-}0 & \phantom{-}1 & \phantom{-}0 & \phantom{-}0\\
   \phantom{-}0 & \phantom{-}0 & \phantom{-}0 &          - 1\\
   \phantom{-}0 & \phantom{-}0 &          - 1 & \phantom{-}0\\
  \epm \ .
\ee

\begin{figure*}
 \includegraphics[width=0.49\textwidth]{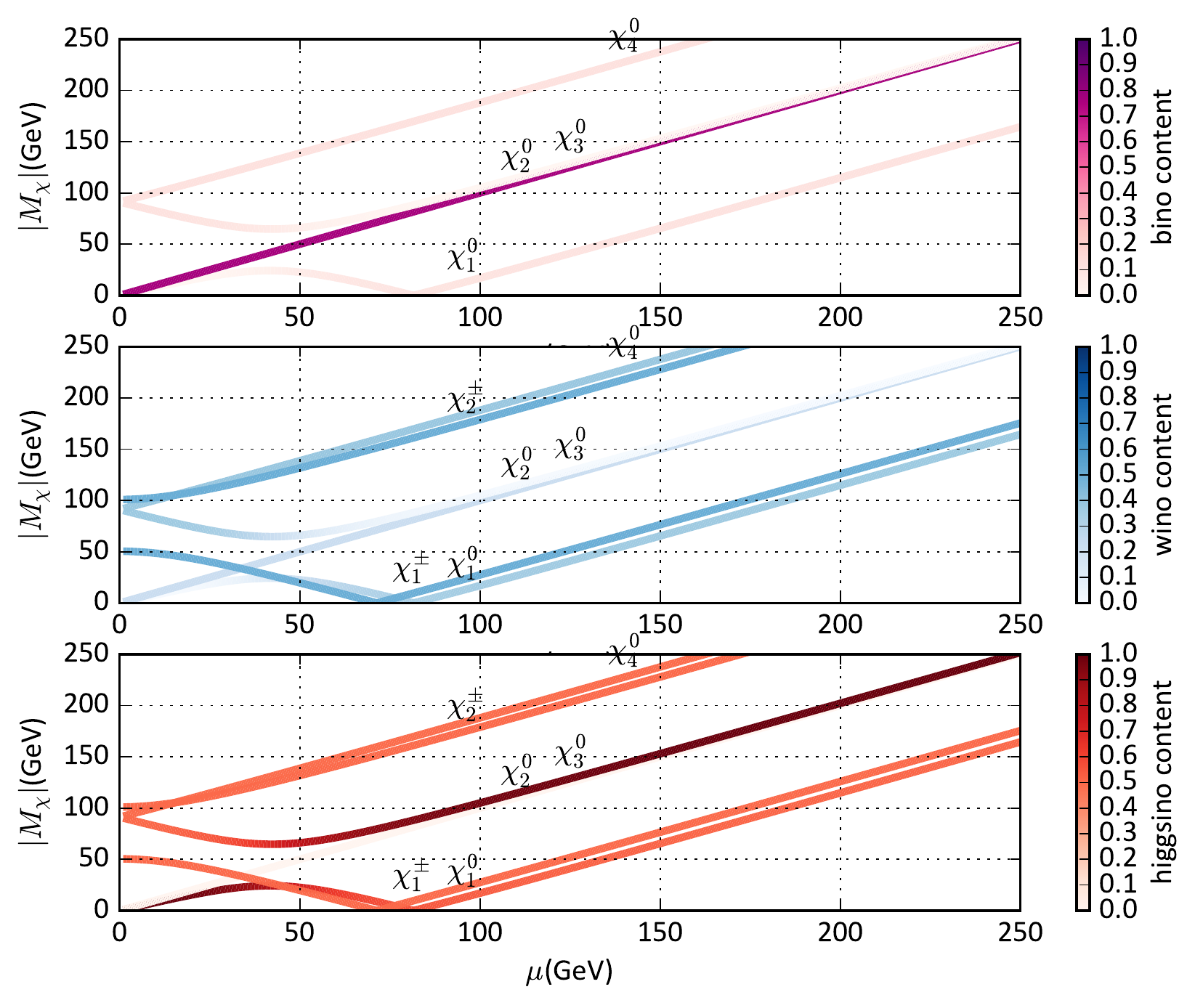}
 \includegraphics[width=0.49\textwidth]{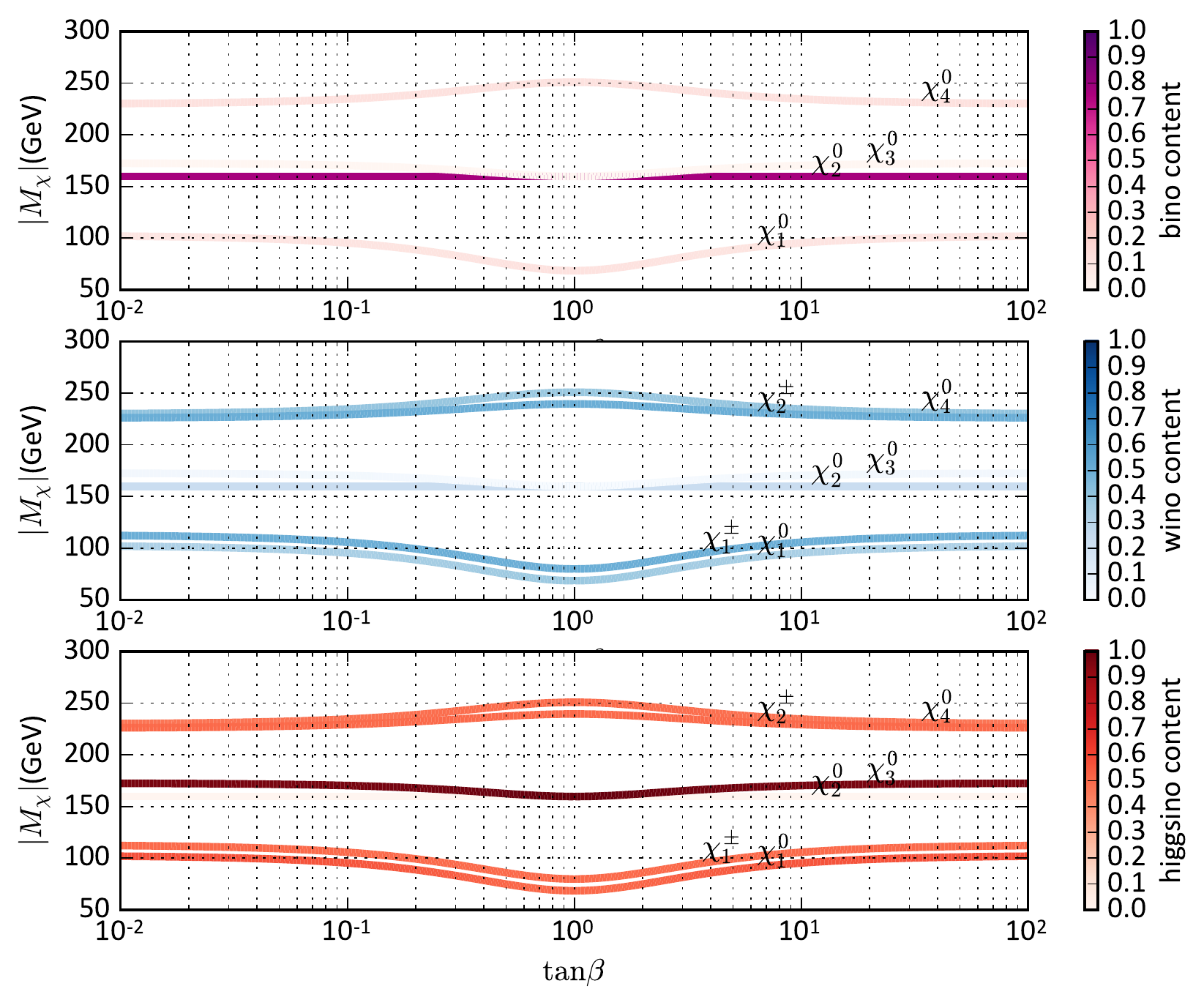}\\
 \includegraphics[width=0.49\textwidth]{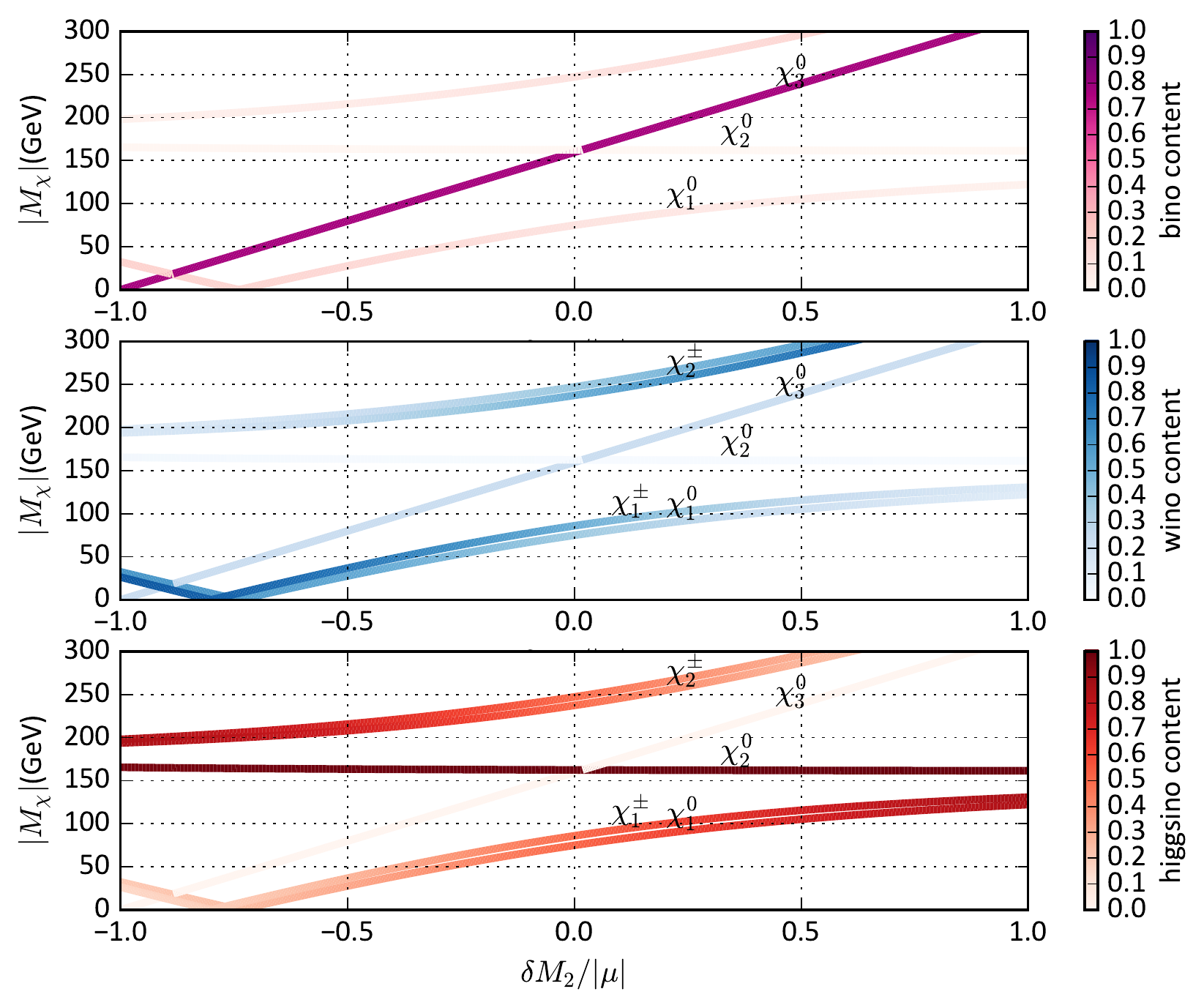}
 \includegraphics[width=0.49\textwidth]{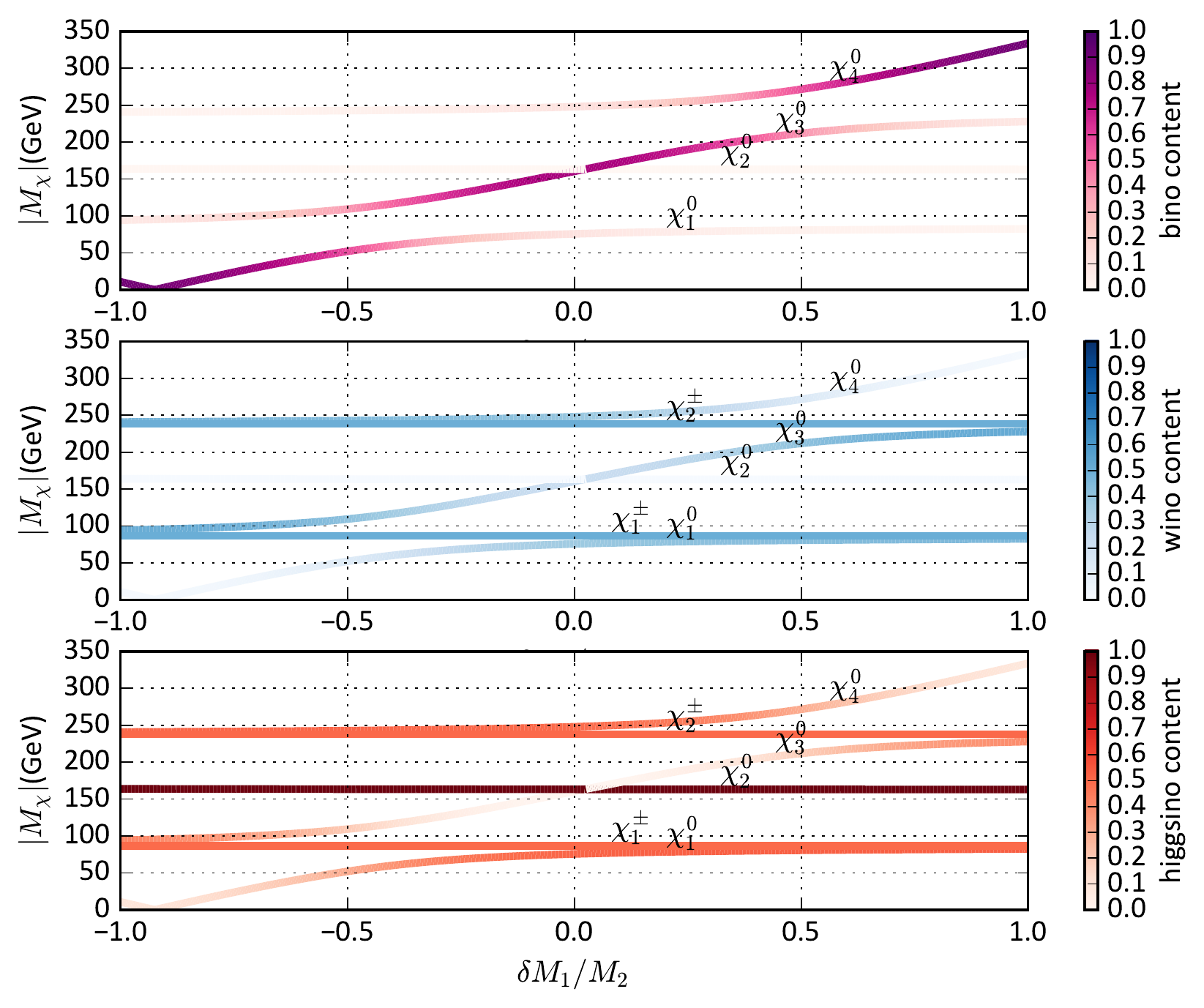}\\
 \caption{\label{fig:mupos} Variation of the neutralino and chargino mass
   spectra for scenarios featuring $\mu > 0$, as a function of $|\mu|$ (upper
   left), $\tan \beta$ (upper right), $\delta M_{2}/|\mu|$ (lower left) and
   $\delta M_{1} / M_{2}$ (lower right) when all other parameters are fixed to
   their reference value given in Eq.~\eqref{eq:default}. The colour-coding (with
   increased line width for better visibility) indicates the bino (purple),
   wino (blue) and higgsino (red) content of the different particles.}
\end{figure*}

\subsection{Mass spectra and gaugino-higgsino content} \label{ssec:2.3}

As stated at the beginning of this section, the parameter space describing the
MSSM gaugino-higgsino sector is four-dimensional and specified by the parameters $\mu$,
$\tan\beta$, $M_{1}$ and $M_2$. For convenience, we trade the gaugino
mass parameters $M_1$ and $M_2$ for the relative mass differences
$\delta M_{2}/|\mu|$ and $\delta M_{1} / M_{2}$ defined by
\be
 M_{2} = |\mu|\ \lb(1+{\delta M_{2}\over|\mu|}\rb)
 \quad {\rm and} \quad
 M_{1} = M_{2}\,\lb(1+{\delta M_{1}\over M_{2}}\rb)\ .
\ee
The resulting mass spectrum and neutralino and chargino decompositions are
related to these parameters in a complex and non-trivial manner, which makes it
difficult to get a global understanding of the response of the spectrum
to a variation in these parameters. Therefore, we explore the parameter space in
a systematic way by first defining a default scenario
\beq
  |\mu| = 2 M_{W}\ ,\ \
  \tan\beta=2 \vee \frac{1}{2}\ ,\ \
  {\delta M_{2}\over|\mu|} = 0\ \ \ {\rm and}\ \
  {\delta M_{1} \over M_{2}} = 0 \ ,
\label{eq:default}\eeq
and then varying one of these parameters at a time.

\begin{figure*}
 \includegraphics[width=0.485\textwidth]{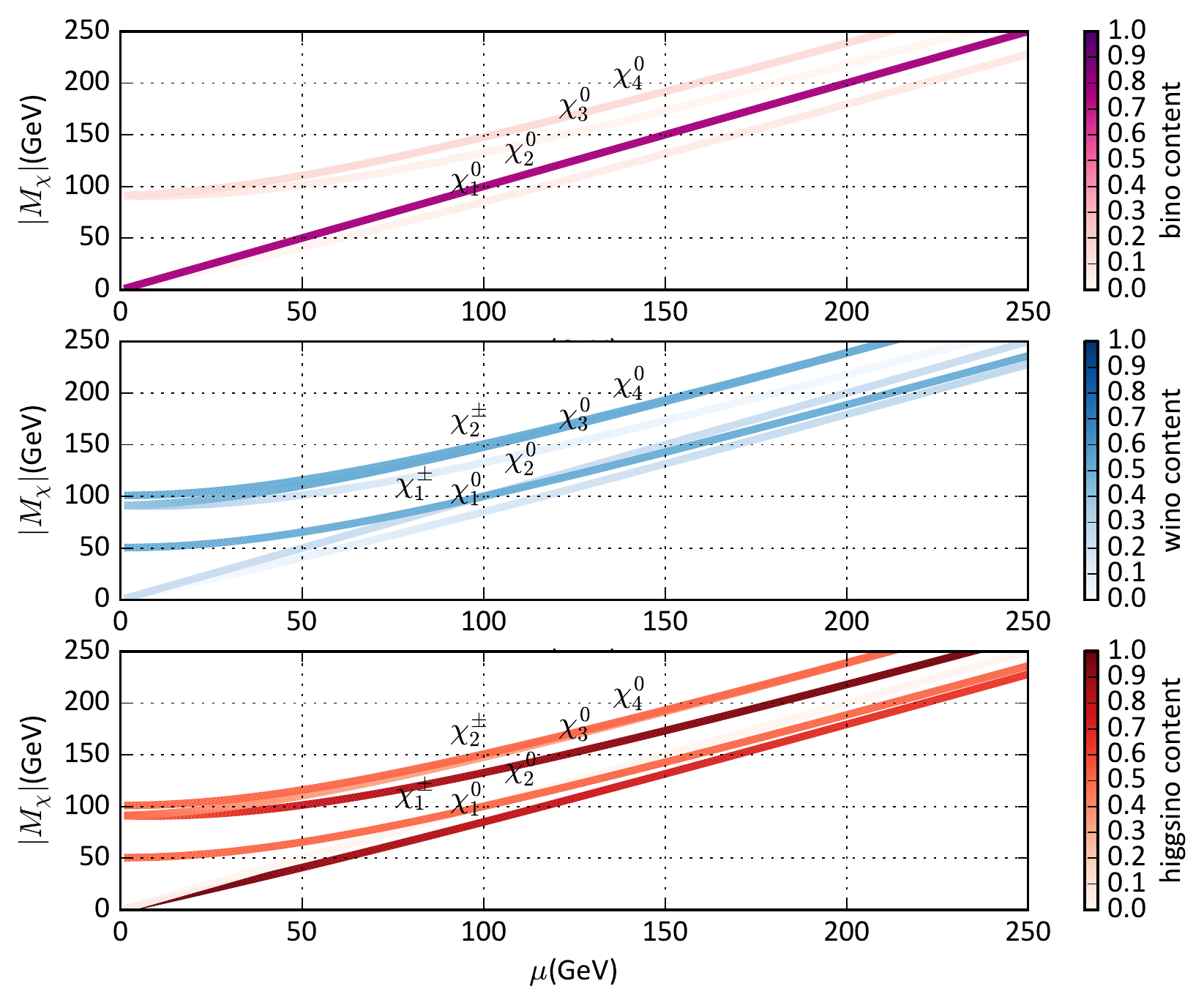}
 \includegraphics[width=0.485\textwidth]{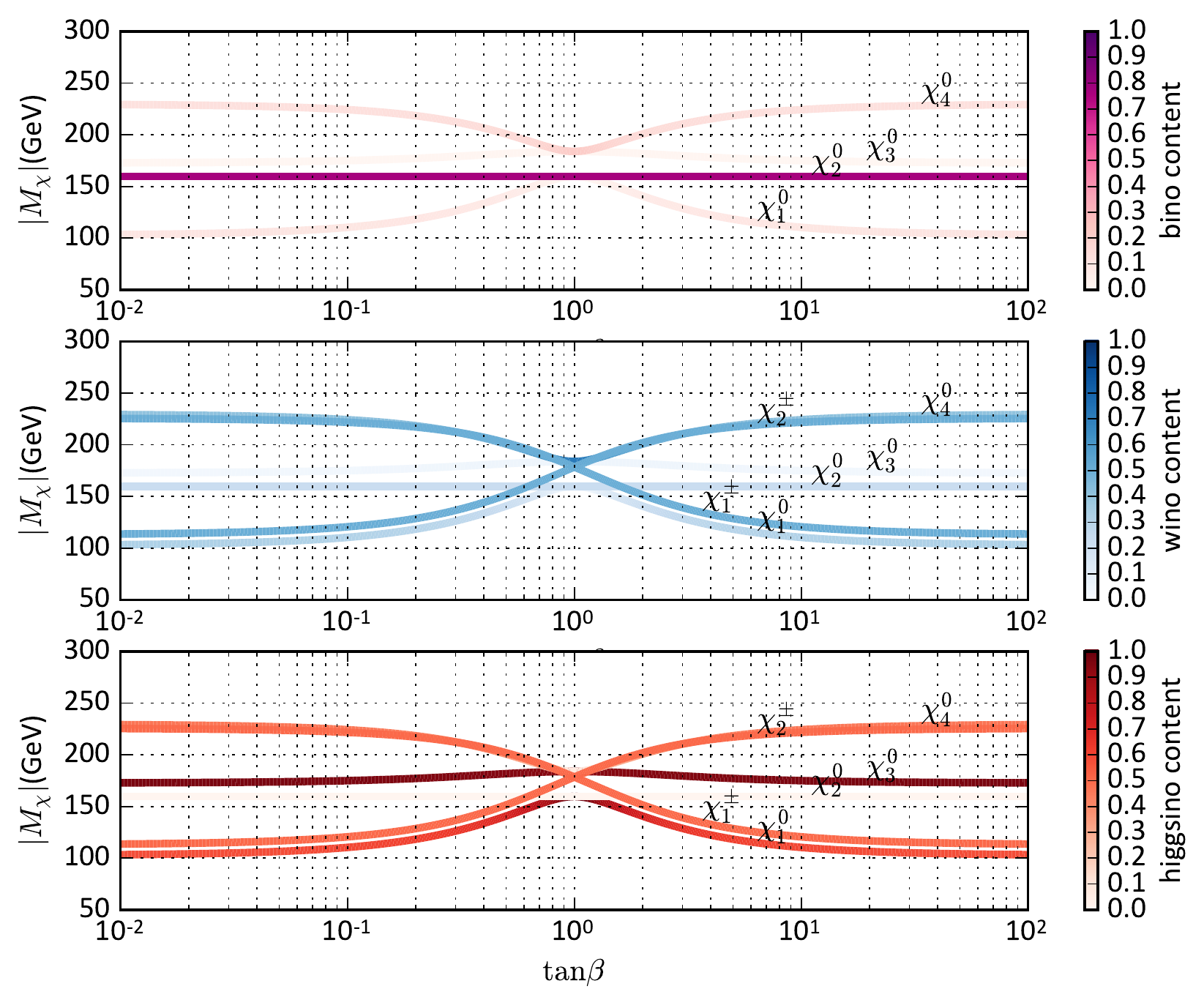}\\
 \includegraphics[width=0.49\textwidth]{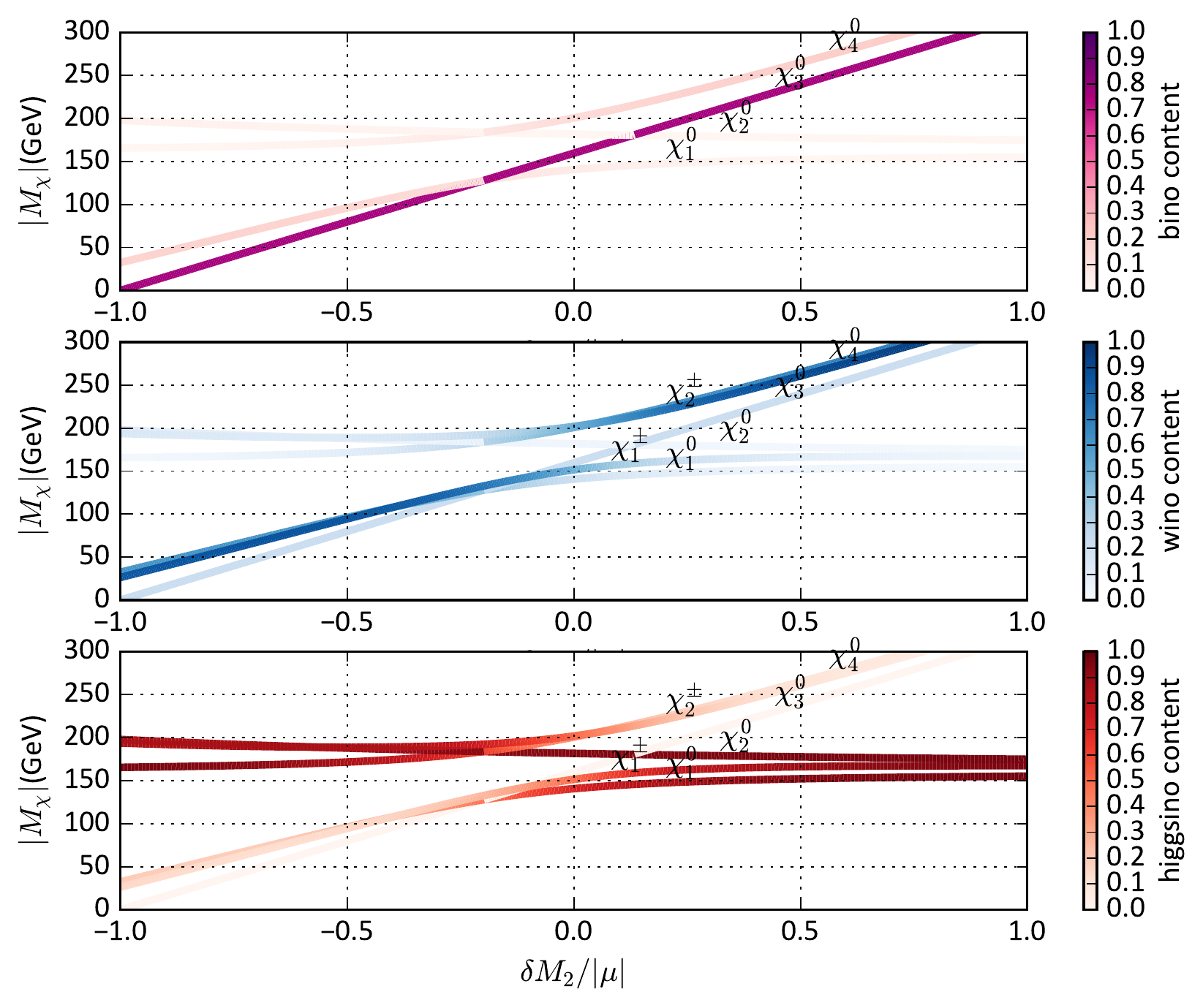}
 \includegraphics[width=0.49\textwidth]{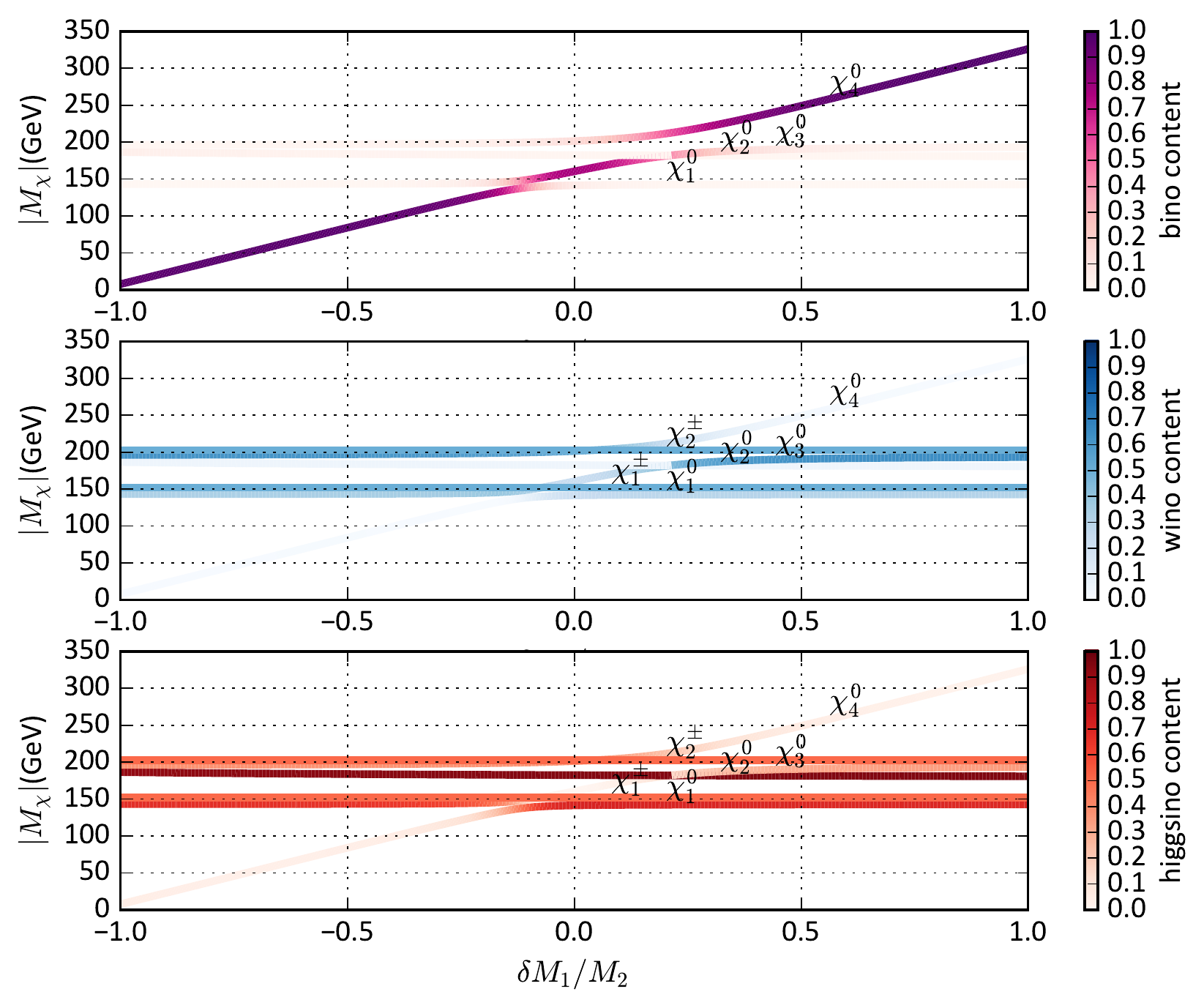}\\
 \caption{\label{fig:muneg} Same as Fig.\ \ref{fig:mupos} for $\mu < 0$.}
\end{figure*}

The results are shown in Figs.~\ref{fig:mupos} and \ref{fig:muneg} for scenarios
featuring a positive and a negative $\mu$ parameter, respectively. In these
figures, we provide a global overview on how a variation of one of the model
input parameters affects the mass spectra and the neutralino and chargino
decompositions in terms of the gaugino and higgsino states. Starting from the
reference scenario of Eq.~\eqref{eq:default}, we vary either the $\mu$ parameter
(upper left panels of the figures), $\tan\beta$ (upper right panels of the
figures), the ratio $\delta M_{2}/|\mu|$ (lower left panels of the figures) or
the ratio $\delta M_{1} / M_2$ (lower right panels of the figures). Although
opposite choices for the sign of $\mu$ correspond to different regions in the
parameter space, they can potentially lead to similar mass spectra ({\it cf.}
the discussion in Sec.~\ref{ssec:2.2}). In the upper, middle and lower parts of
each subfigure, we show the respective dependence of the bino (only for
neutralinos), wino and higgsino content of each electroweakino state on the
considered model parameter.
%
Trivially, we retrieve the fact that the chargino sector does not depend on the
bino mass parameter $M_1$, and thus also not on $\delta M_{1} / M_{2}$.

The mixing pattern of the gaugino and higgsino states is driven by the
off-diagonal elements in the mass matrices of Eq.~\eqref{eq:chargino-1} and
Eq.~\eqref{eq:neutralino-1}, which are all roughly proportional to the $W$-boson
mass. Therefore, maximally mixed states arise only when either $|\mu|$, $M_1$,
$M_2$, $|\pm\mu-M_{1}|$, $|\pm\mu-M_{2}|$ or $|M_{1}-M_{2}|$ is of
$\mathcal{O}(M_W)$ or smaller. Conversely, nearly pure gaugino and higgsino
states in the chargino sector occur for $|\mu|\gtrsim M_{W}$ and
$|\mu-M_{2}|\gtrsim M_W$, while pure states in the neutralino sector
additionally require also $|-\mu-M_{2}|\gtrsim M_W$ and $|\pm\mu-M_{1}|\gtrsim
M_W$.

The diagonalisation of the chargino and neutralino mass matrices can possibly
yield negative mass eigenvalues. In this case, they are made positive by
absorbing the sign into the mixing matrices that get imaginary, which thus
affects the couplings. The original sign of the mass can be deduced by examining
the variation of the curves in Fig.~\ref{fig:mupos} and Fig.~\ref{fig:muneg}. A
change of sign can be traced back to a curve hitting zero and exhibiting a
discontinuous local derivative. This configuration occurs when any of the
$\mu$, $M_{1}$ and $M_{2}$ mass parameters is of $\mathcal{O}\lb(M_{W}\rb)$.
Outside this range, two of the neutralinos always feature a dominant higgsino
content, and their masses have opposite signs. The sign of the masses of the
other two neutralinos is driven by the sign of the $M_{1}$ or $M_{2}$
parameters, depending on the dominant bino or wino nature of the neutralinos
under consideration, provided the mixing is small. Moreover, one observes that
the neutralino mass lines can only cross if the masses have opposite signs.
Otherwise, one gets an avoided crossing where the neutralino content is
exchanged.

The results presented in the upper left panels of Fig.~\ref{fig:mupos} and
Fig.~\ref{fig:muneg} confirm that when $|\mu|$, and subsequently also $M_{1}$
and $M_{2}$, exceeds the $W$-boson mass scale, the overall magnitude of the
electroweakino masses is solely set by $|\mu|$ and increases uniformly with it.
In the special case corresponding to $\delta M_{2}/|\mu|=\delta M_{1}/M_{2}=0$,
the mass differences as well as the elecroweakino decompositions moreover become
independent of $|\mu|$.
%
In contrast, variations of $\delta M_{2} / |\mu|$ and $\delta M_{1} / M_{2}$
influence the electroweakino mass differences, as shown in the lower panels of
Fig.~\ref{fig:mupos} and Fig.~\ref{fig:muneg}. These parameters are thus those
that will allow us to determine MSSM benchmark points defined by an overall mass
scale and a given mass splitting between the superpartners. In particular, one
can obtain a spectrum where the lighter (heavier) states are nearly pure
higgsinos when $\delta M_{2}/|\mu| \gg 0$ ($\delta M_{2}/|\mu| \ll 0$).
Different values of $\tan \beta$ or $\delta M_{1}/M_{2}$ then raise or lower the
value at which the turnover occurs. Similarly, varying both $\delta M_2/|\mu|$
and $\delta M_{1} / M_{2}$ allows one to obtain scenarios featuring nearly pure
bino or wino states as the heaviest or lightest states.
%
Finally, as illustrated on the upper right panels of the figures, we observe
that the $\tan\beta$-dependence of the spectrum exhibits a peak or a dip at
$\tan\beta=1$ with an amplitude that is typically smaller than about $M_W/2$.
Except this feature, the effect of $\tan \beta$ on the spectrum is small, which
therefore allows us to use this parameter for small adjustments once all other
parameters have been chosen. When $\tan \beta \ll 1$ or $\tan \beta \gg 1$, the
dependence on $\tan\beta$ moreover vanishes.

The sign of the $\mu$ parameter has little influence on the mass spectrum upon
variations of $|\mu|$, $\delta M_{2}/|\mu|$ or $\delta M_{1} / M_{2}$. Negative
$\mu$ values only induce a more compressed spectrum compared to the case of a
positive $\mu$ parameter. In the chargino sector, the opposite signs of $\mu$
and $M_{2}$ in this case lead to an unavoided crossing of the mass eigenvalues
at $\tan \beta \sim 1$, as well as to an opposite behavior when increasing or
decreasing $\tan\beta$ with respect to 1. For $\mu < 0$, gaugino-higgsino
mixings and eletroweakino mass splittings indeed increase with $\tan\beta$
variations, whilst they decrease for $\mu > 0$.
%
In addition, mixed bino/wino-states are rare and can only be obtained by
fine-tuning the parameters due to the non-existence of any direct bino/wino
coupling in the Lagrangian. Moreover, wino states mix more easily with higgsino
states than bino states as the hypercharge and weak couplings satisfy $g_{Y} < g_{2}$,
or equivalently as $\sin\theta_{W}<\cos \theta_{W}$.

\section{Scan strategy} \label{sec:3}

In this section, we first select a strategy to explore the parameter space
of the MSSM gaugino/higgsino sector in an efficient way. We then define
criteria for acceptable benchmark points that fit best a pre-defined
mass spectrum of light neutralinos and charginos and discuss how additional
requirements, such as a large higgsino content of these sparticles, can
also be included. Finally, we briefly reflect on possible generalisations
of these strategies.

\subsection{Parameter space exploration} \label{ssec:3.1}

The observations made in the previous section allow for the identification of
general characteristics of the gaugino-higgsino parameter space that are useful
for building realistic benchmark scenarios. Following most of the experimental
studies at the LHC, we focus on configurations with only two light neutralinos
($\chi^{0}_{1}$, $\chi^{0}_{2}$) and a light chargino ($\chi^{\pm}_{1}$).
For illustrative purposes, we take the desired chargino mass $M_{\chi^\pm_1}$
as an input and ask for an equidistant mass splitting $\Delta M_{21}$
of the two neutralinos,
\be \label{eq:modelscan}
  M_{\chi^0_1}   = M_{\chi^\pm_1} - \frac{\Delta M_{21}}{2}
     \qquad\text{and}\qquad
  M_{\chi^0_2}   = M_{\chi^\pm_1} + \frac{\Delta M_{21}}{2} \ .
\ee
The scan procedure described in the following can, however, easily be generalised to
other setups.

In principle, we scan over all four parameters $\mu$, $\tan \beta$, $M_{1}$ and
$M_{2}$, but we immediately reduce this parameter space on the basis of the
transformations that leave the neutralino-chargino mass spectrum invariant.
Regions of the parameter space that are not explored are then derived by
transforming the mixing matrices as described in Sec.~\ref{ssec:2.2}. As a
consequence, we only scan over the regions $\tan\beta\in [1;100]$ and $M_{2}>0$.
In contrast, the sign of the higgsino mass parameter $\mu$ can strongly affect
the structure of the theory and the experimental signatures, so that we consider
both $\mu < 0$ and $\mu > 0$. As upper bounds on the absolute values of the
mass parameters, we impose 5 TeV, which we only raise when we see that our results
cluster near them. Values of $M_2<0$ and $\tan\beta<1$ are obtained with sign
flips and a $\tan\beta$ inversion, as explained in Sec.~\ref{ssec:2.2}.
The three dimensionful parameters $\mu$, $M_1$ and $M_2$ are finally further
constrained by the requirements on the desired gaugino/higgsino decomposition.

As an illustration of the above strategy, we search for benchmark scenarios
featuring a spectrum where the lightest states are all higgsino-like. The range
of $\mu$ can then be restricted by observing ({\it cf.} Sec.~\ref{ssec:2.3}) that the
masses of neutralinos and charginos with a dominant higgsino contribution lie
in the range $|\mu| \pm \mathcal{O}\lb(M_{W}\rb)$. The scan ranges are thus given by
\be\label{eq:scanranges-1}\bsp
  |\mu| \in&\ \Big[ \min(M_{\chi^\pm_1}) \!-\! \mathcal{O}(M_W), \
       \max(M_{\chi^\pm_1})\!+\!\mathcal{O}(M_W) \Big] \ ,\\
  |M_{1}| \in&\ \Big[ \min(M_{\chi^\pm_1})\!-\!\mathcal{O}(M_W),\ 5~{\rm TeV}\Big]\ ,\\
  M_2 \in&\ \Big[ \min(M_{\chi^\pm_1})\!-\!\mathcal{O}(M_W),\ 5~{\rm TeV}\Big] \ ,\\
  \tan\beta \in&\ \lb[ 1, 100 \rb]\ ,
\esp\ee
with
\be
  \text{sign}\lb(\mu\rb) \in \{-,+\} \qquad\text{and}\qquad
  \text{sign}\lb(M_{1}\rb) \in \{-,+\}\  ,
\ee
and where $\min(M_{\chi^\pm_1})$ and $\max(M_{\chi^\pm_1})$ represent the minimal and maximal
desired values for the light chargino mass.

It is easy to see that an equidistant scan in these parameters is not very efficient.
For instance, variations at large values of $\tan \beta$ only weakly
affect the spectrum and the gaugino/higgsino decompositions, since these depend on
$\sin \beta$ and $\cos \beta$ rather than $\tan \beta$. Also, a scan over
multiple orders of magnitude for the gaugino mass parameters does not
efficiently cover masses in the lower ranges where $|M_{1}|,|M_{2}|\sim |\mu|$,
where the masses of the higgsino-like neutralinos and charginos are affected the
most. We therefore reparameterise the prior distributions in $M_1$, $M_2$ and
$\tan\beta$ as
\be\label{eq:scanranges-2}\bsp
  M_1 =&\ \pm  M \lb(1 - \frac{\epsilon}{2} \rb)\ ,\\
  M_2 =&\      M \lb(1 + \frac{\epsilon}{2} \rb)\ , \\
  \tan\beta = &\ \tan\beta_{\min}
    \exp\bigg[x_\beta \ln\frac{\tan\beta_{\max}}{\tan\beta_{\min}}\bigg]
\esp\ee
with
\be\label{eq:scanranges-2b}\bsp
  &M = M_{\min} \exp\bigg[ x_M\ln\frac{M_{\max}}{M_{\min}}\bigg] \ ,\\
  &x_{M}     \in \lb[0,1\rb]\ , \\
  &\epsilon \in \lb[-2,2\rb]\ , \\
  &x_{\beta} \in \lb[0,1\rb]\ .
\esp\ee
In the expressions of Eq.~\eqref{eq:scanranges-2} and
Eq.~\eqref{eq:scanranges-2b}, the minimum and maximum values of $\tan\beta$ and $M$
are dictated by the scan range, with $M_{\min}$ and $M_{\max}$
referring both to $M_1$ and $M_2$.
%
The scan time can be further reduced with an iterative procedure, where at each
iteration the parameter range in $|\mu|$, $x_{M}$, $\epsilon$, and $x_{\beta}$ is
halved keeping the currently best parameters central. The total parameter space
volume then shrinks each time by a factor of $(1/2)^4=1/16$ with an additional
factor of 1/2 from the sign determination of $M_{1}$ in the first iteration.

\subsection{Benchmark selection} \label{ssec:3.2}

The quality of our fit of the desired mass spectrum is parameterised by
the relative differences between the input masses and their fit values
compared to the corresponding grid spacings $\Delta M_{\chi^{\pm}_{1}}$ and
$\Delta(\Delta M_{21})$,
\be\label{eq:demands}\bsp
  d_{1} =&\ \frac{1}{\Delta M_{\chi^{\pm}_{1}}}
    \bigg[M_{\chi^\pm_1}-M_{\chi^\pm_1}^{\rm fit}\bigg] =:
    \frac{\delta M_{\chi^{\pm}_{1}}}{\Delta M_{\chi^{\pm}_{1}}}\ , \\
  d_{2} =&\ \frac{2}{\Delta(\Delta M_{21})}
    \bigg[\frac{\Delta M_{21}}{2}\!-\!
      \lb(M_{\chi^0_2}^{\rm fit}\!-\!M_{\chi^\pm_1}^{\rm fit}\rb)\bigg]\!=:\!
    \frac{\delta (\Delta M_{21})}{\Delta(\Delta M_{21})}\ , \\
  d_{3} =&\ \frac{2}{\Delta(\Delta M_{21})}
    \bigg[\frac{\Delta M_{21}}{2}\!-\!
      \lb(M_{\chi^\pm_1}^{\rm fit}\!-\!M_{\chi^0_1}^{\rm fit}\rb)\bigg]\!=:\!
    \frac{\delta' (\Delta M_{21})}{\Delta(\Delta M_{21})}\ .
\esp\ee
A perfect fit then has $d_{1}=d_{2}=d_{3}=0$, while the penalty score of a
configuration with respect to its nearest neighbour grid point is given by
\be\label{eq:defscore} 
  \text{score} = \sqrt{\frac{d^{2}_{1}+d^{2}_{2}+d^{2}_{3}}{3}}
\ee
%
with $\sqrt{1/3}\sim 0.58$ for a nearest neighbour grid point with a single
outlier. We consider a configuration acceptable if
\beq
 {\rm score}<0.1,
\eeq
which represents a reasonable compromise between scan time
and accuracy:
\be
 \label{eq:acc_config} \bsp
  & \delta M_{\chi^{\pm}_{1}} ~<~ 0.1 \sqrt{3} \Delta M_{\chi^{\pm}_{1}} ~=~
     0.17 \Delta M_{\chi^{\pm}_{1}} \ , \\
  & \frac{\delta^{(')} (\Delta M_{21})}{2} ~<~ 0.1 \sqrt{3} \Delta (\Delta M_{21}) ~=~
     0.17 \Delta(\Delta M_{21}) \ .
\esp\ee
While this procedure allows us to find an approximately correct chargino and neutralino
mass spectrum, it still does not maximise their average higgsino (or gaugino) content.
This type of additional condition can be included by reweighting the score with
\be\label{eq:max_config}
  \text{score}_{\text{new}} \
    \frac{\tilde{f}_{\text{old}}}{\widetilde{f}_{\text{new}}} <
    \text{score}_{\text{old}} \ ,
\ee
which balances accuracy of mass spectrum and decomposition for scores
that are neither too small nor too large. In the case study of Sec.\ \ref{sec:4},
$\tilde f$ represents the average higgsino content of the light neutralinos
$\chi^0_1$, $\chi^{0}_{2}$ and the light $\chi^{\pm}_{1}$ particles.

\subsection{Generalisation} \label{ssec:3.3}

The specific setup described above can be generalised by modifying the desired
mass spectrum of Eq.~\eqref{eq:modelscan} to non-equidistant mass differences
with the according adjustments in the conditions of Eq.\eqref{eq:demands}.
A qualitatively very distinct modification is the requirement of one-sided mass limits.
Second, the maximisation of the higgsino content through the function $\tilde f$
and/or the reweighting condition in Eq.\ \eqref{eq:max_config} can be replaced.
A specific example would be the maximisation of couplings to specific particles.
In practice, a trial scan often helps in defining more precisely acceptable
configurations and conditions that do not overly constrain the interesting
regions of parameter space. Scan ranges can often be guessed by using the
observations made in Sec.~\ref{ssec:2.3}. Reparameterisations as the one in
Eq.~\eqref{eq:scanranges-2} are moreover useful when scanning over multiple orders
of magnitude in one or several parameters and can be optimised by studying the
posterior distributions in the input parameters.

\section{Case study: Higgsino-like neutralinos and charginos} \label{sec:4}

\begin{table}
 \caption{\label{tab:1}Targeted mass ranges, splittings and spacings for light higgsino-like neutralinos and charginos.}
 \begin{center}
  \begin{tabular}{|c|rrr|}
   \hline
    Mass / splitting  & Minimum  & Maximum  & Grid spacing \\
    \hline
    $M_{\chi^{\pm}_{1}}$ & $90$ GeV & $400$ GeV & $3.1$ GeV \\ 
    $\Delta M_{21}$         & $1$ GeV   & $100$ GeV & 1 GeV\\
    \hline
  \end{tabular}
 \end{center}
\end{table}
\begin{figure*}
   \begin{center} 
   \subfigure[score ($\mu > 0$)]{\includegraphics[width=0.48\textwidth]{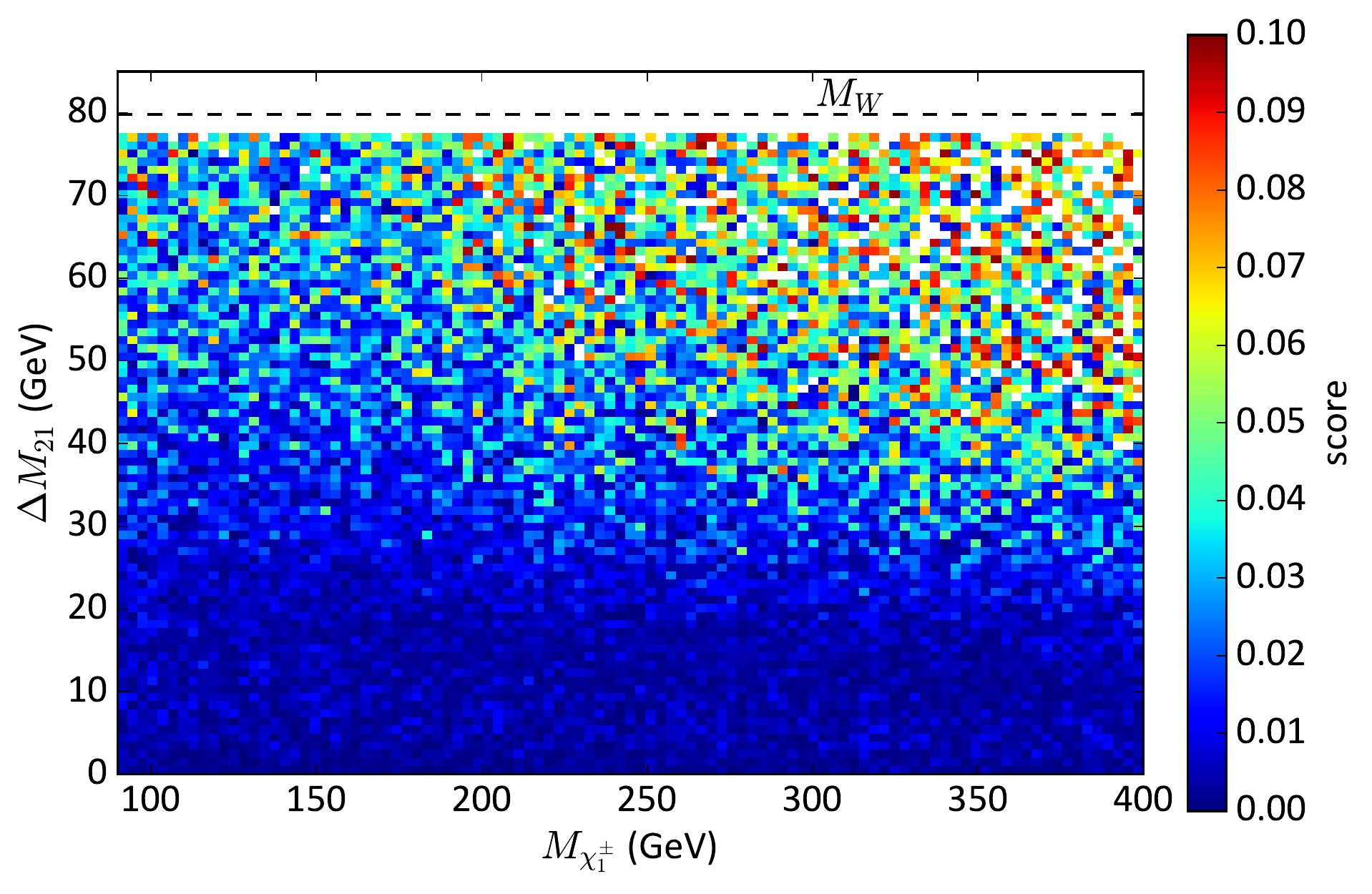}}
   \subfigure[score ($\mu < 0$)]{\includegraphics[width=0.48\textwidth]{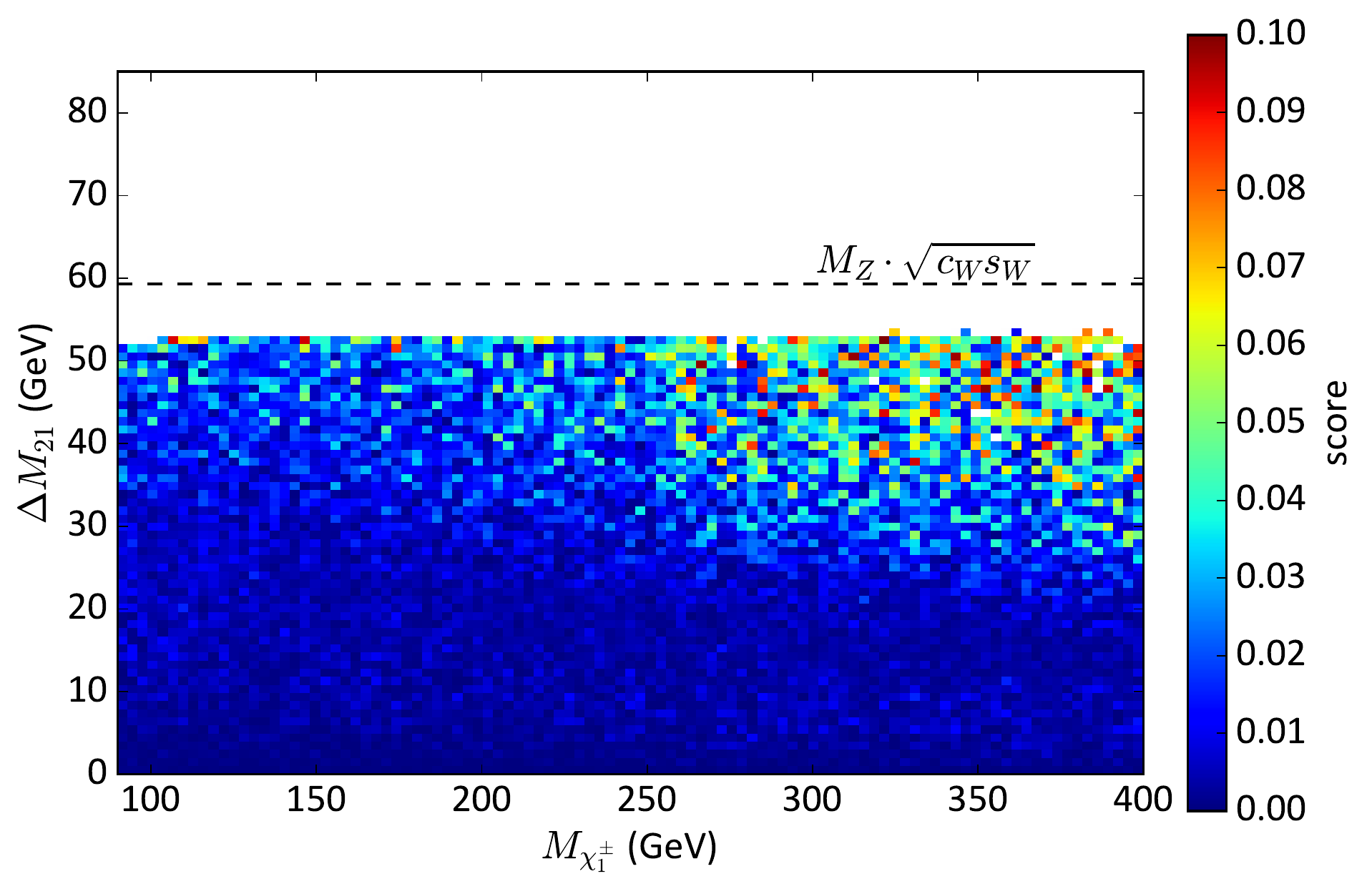}} \\[-3mm]
   \subfigure[higgsino content ($\mu > 0$)]{\includegraphics[width=0.48\textwidth]{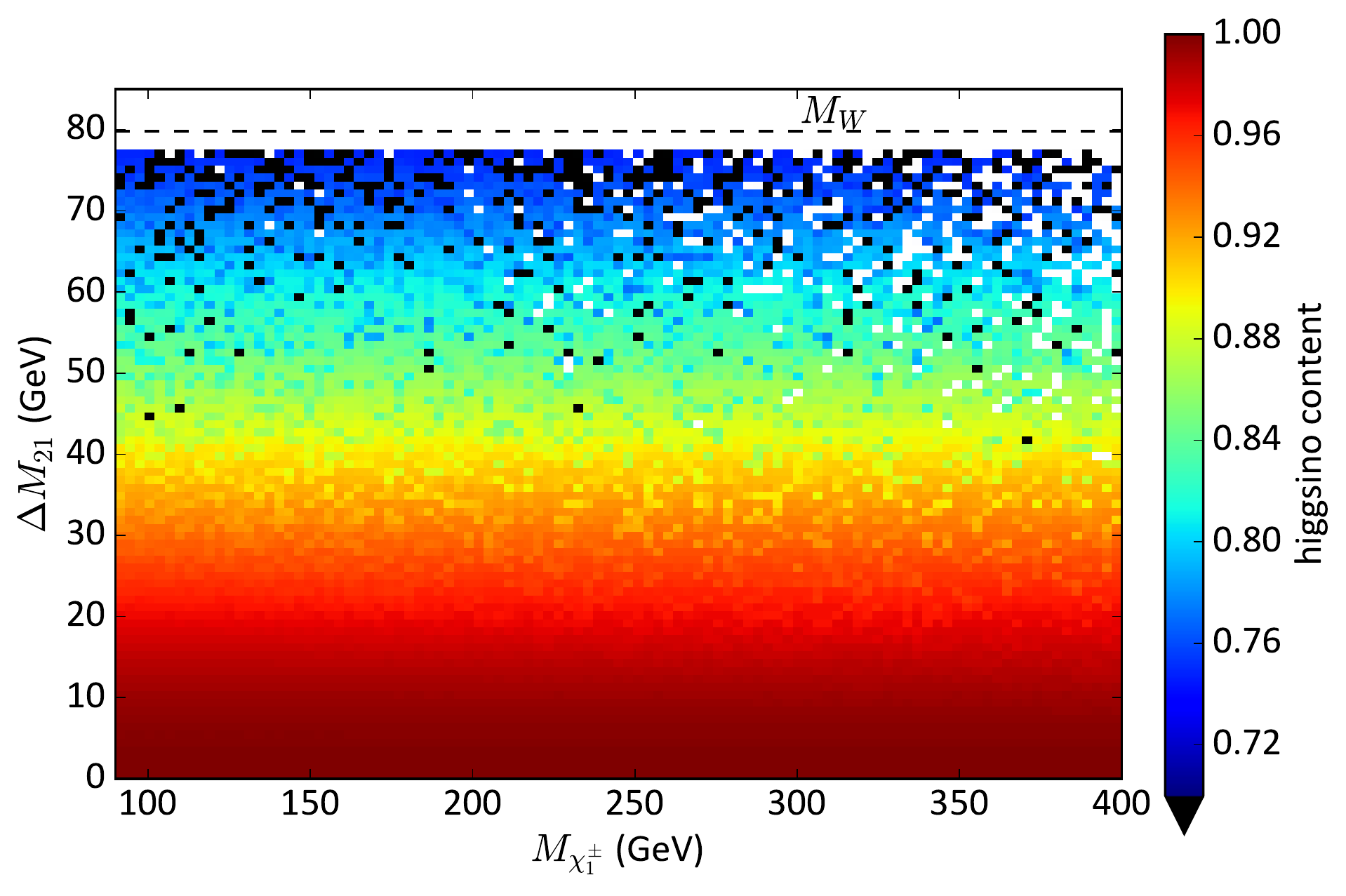}}
   \subfigure[higgsino content ($\mu < 0$)]{\includegraphics[width=0.48\textwidth]{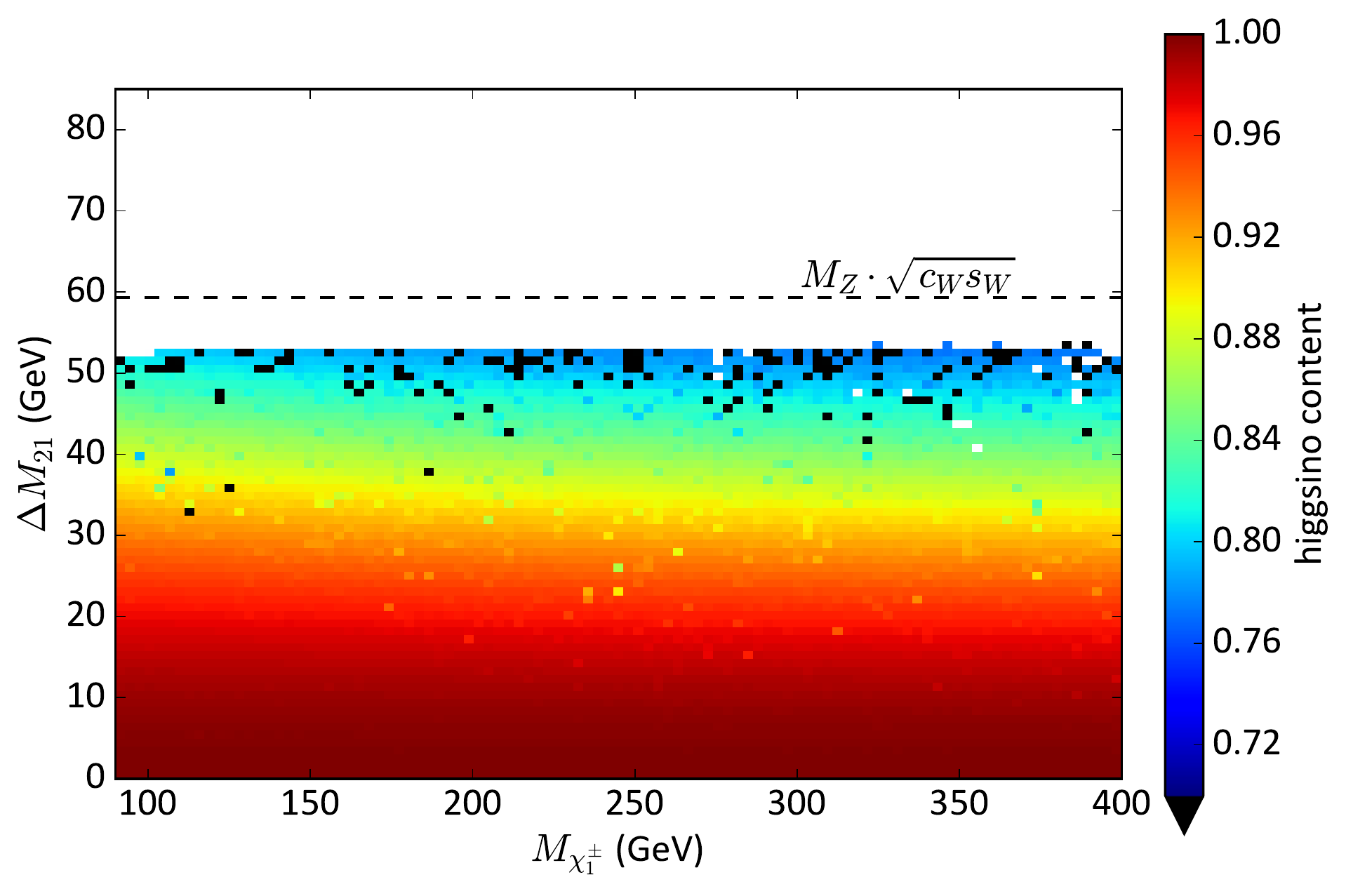}}
    \end{center}
 \caption{\label{fig:firsteval}Scores and average higgsino contents of $\chi^{0}_{1}$,
 $\chi^{0}_{2}$ and $\chi^{\pm}_{1}$ for the models found in our MSSM fits of light chargino mass
 $M_{\chi^\pm_1}$ and neutralino mass splitting $\Delta M_{21}$. The dashed lines indicate that the
 latter are always smaller than $\mathcal{O} \lb(M_{W}\rb)$ (remember that $M_W=M_Zc_W$).}
\end{figure*} 
%
In this section, we present a case study of a specific simplified MSSM model with a realistic
neutralino-chargino sector, whose general properties were discussed in Sec.\ \ref{sec:2}.
We then apply and test the parameter scan method presented in Sec.\ \ref{sec:3} and examine the
properties of the underlying benchmark points. Our case study has higgsino-like light neutralinos
and charginos with equidistant mass splitting and includes both signs of the higgsino mass
parameter $\mu$.

\begin{figure}
   \begin{center} 
   \includegraphics[width=0.45\textwidth]{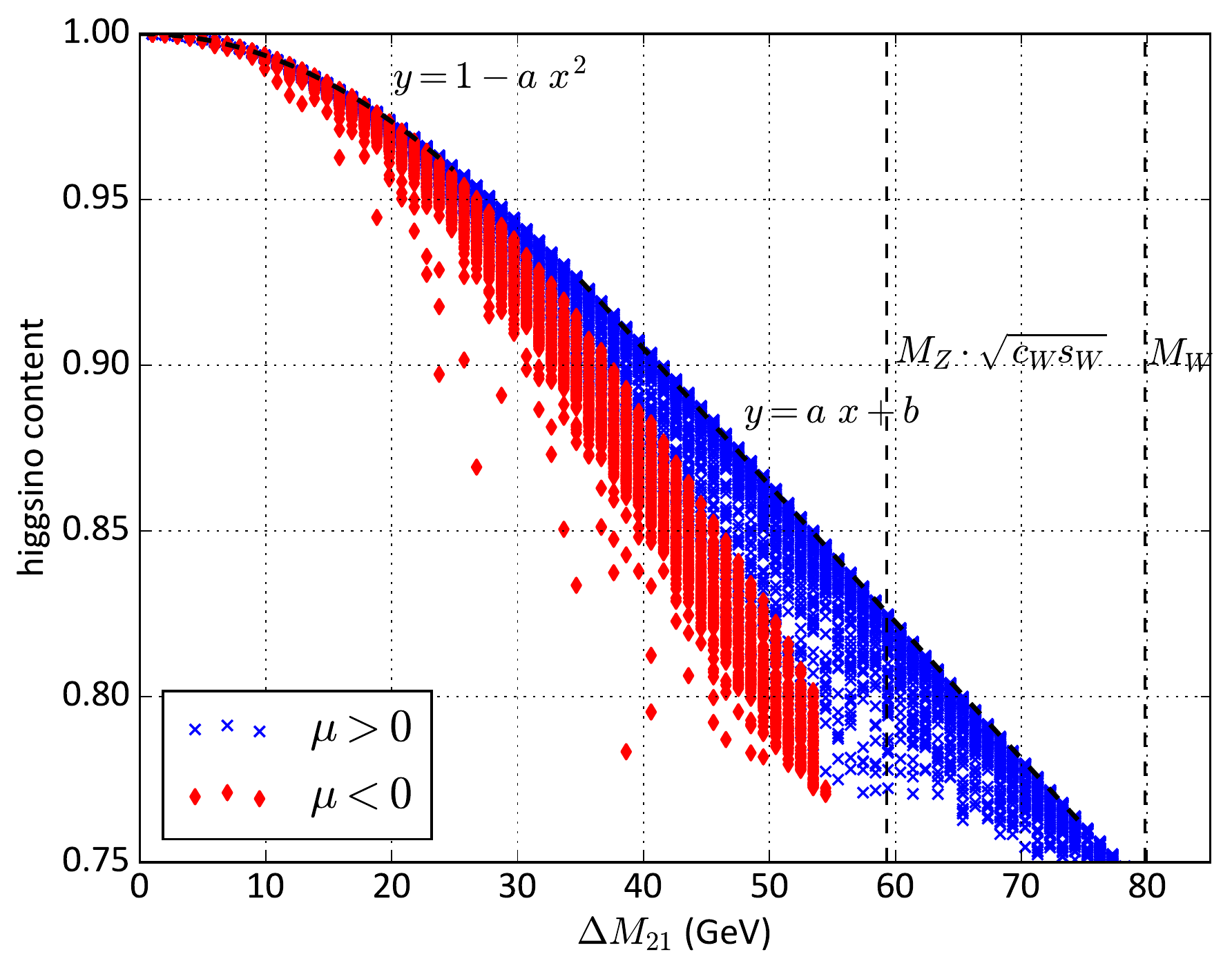}
    \end{center}
 \caption{\label{fig:quanteval}Higgsino content as a function of $\Delta M_{21}$ for our fit
 scenarios with $\mu > 0$ (blue crosses) and $\mu < 0$ (red diamonds). In both cases it
 falls quadratically for $\Delta M_{21} < 25 \ \text{GeV}$. For $\mu>0$, the fall-off is linear
 beyond this value.}
\end{figure}

\subsection{Definition of the simplified model} \label{ssec:4.1}

As it is usually done in simplified models, we decouple the sparticles that are not of
direct relevance to our study, {\it i.e.} squarks, gluinos, and non-SM Higgs particles, by
setting their masses to a sufficiently high value, here $1.5$ TeV. Their phenomenological
impact at the LHC is then negligible due to limited kinematical phase space, suppressed
virtual propagators, and parton distribution functions that vanish at large parton momentum
fractions. Decoupling sparticles with unrealistically high mass values can result
in numerical instabilities in the employed Monte Carlo generators, {\it e.g.} from missing
cancellations in higher-order corrections, and should be avoided.

The targeted light neutralino and chargino mass spectra are defined by a set of
central light chargino masses $M_{\chi^{\pm}_{1}}$ and correlated light neutralino masses
that are split in an equidistant way by $\Delta M_{21}$ ({\it cf.} Eq.\ (\ref{eq:modelscan})).
Their ranges are constrained empirically through negative experimental searches for neutralinos
or charginos, whose masses must exceed the mass of the $Z^{0}$-boson, and theoretically
({\it cf.} Sec.\ \ref{ssec:2.3}) to mass splittings of the two higgsino-like neutralinos that
do not exceed $\mathcal{O}(M_{W})$. We therefore aim to fit the $\mathcal{O}(10^4)$
mass spectra in the ranges shown in Tab.\ \ref{tab:1} by scanning the parameter space as
described in Sec.\ \ref{sec:3}. The lower limit on the chargino mass is inspired
by the combined LEP limit of 92.4 GeV in the higgsino region for any lightest neutralino
mass. This limit rises to 103.5 GeV for mass splittings larger than 5 GeV \cite{lepsusy,%
particledatagroup:2014}.

\subsection{Quality of the scan} \label{ssec:4.2}

The quality of our MSSM fits of these predefined desired scenarios can be evaluated in Fig.~\ref{fig:firsteval}, where we show the distribution of scores defined in Eq.\ (\ref{eq:defscore})
for $\mu > 0$ (upper left) and $\mu<0$ (upper right) as well as the average higgsino contents of
$\chi^{0}_{1}$, $\chi^{0}_{2}$ and $\chi^{\pm}_{1}$ (lower left and right, respectively). The
size of the deviations between the targeted and fitted physical masses can be deduced from the
scores using Eq.\ (\ref{eq:acc_config}).
%

\begin{figure*}
   \begin{center} 
   \subfigure[$|\mu|$ ($\mu > 0$)]
  {\label{fig:mupar-dista} 
   \includegraphics[width=0.46\textwidth]{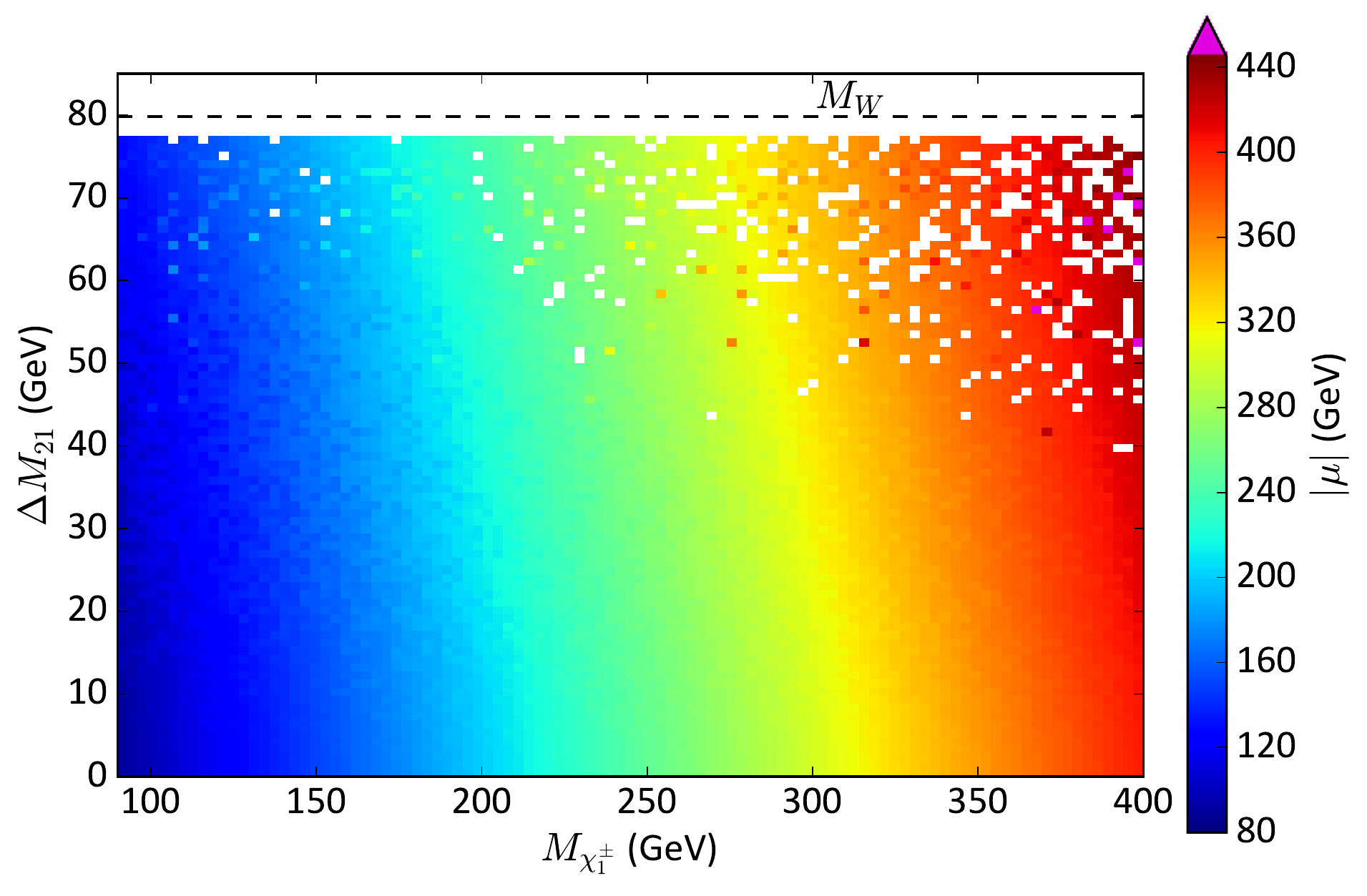}}
   \subfigure[$|\mu|$ ($\mu < 0$)]
  {\label{fig:mupar-distb} 
   \includegraphics[width=0.46\textwidth]{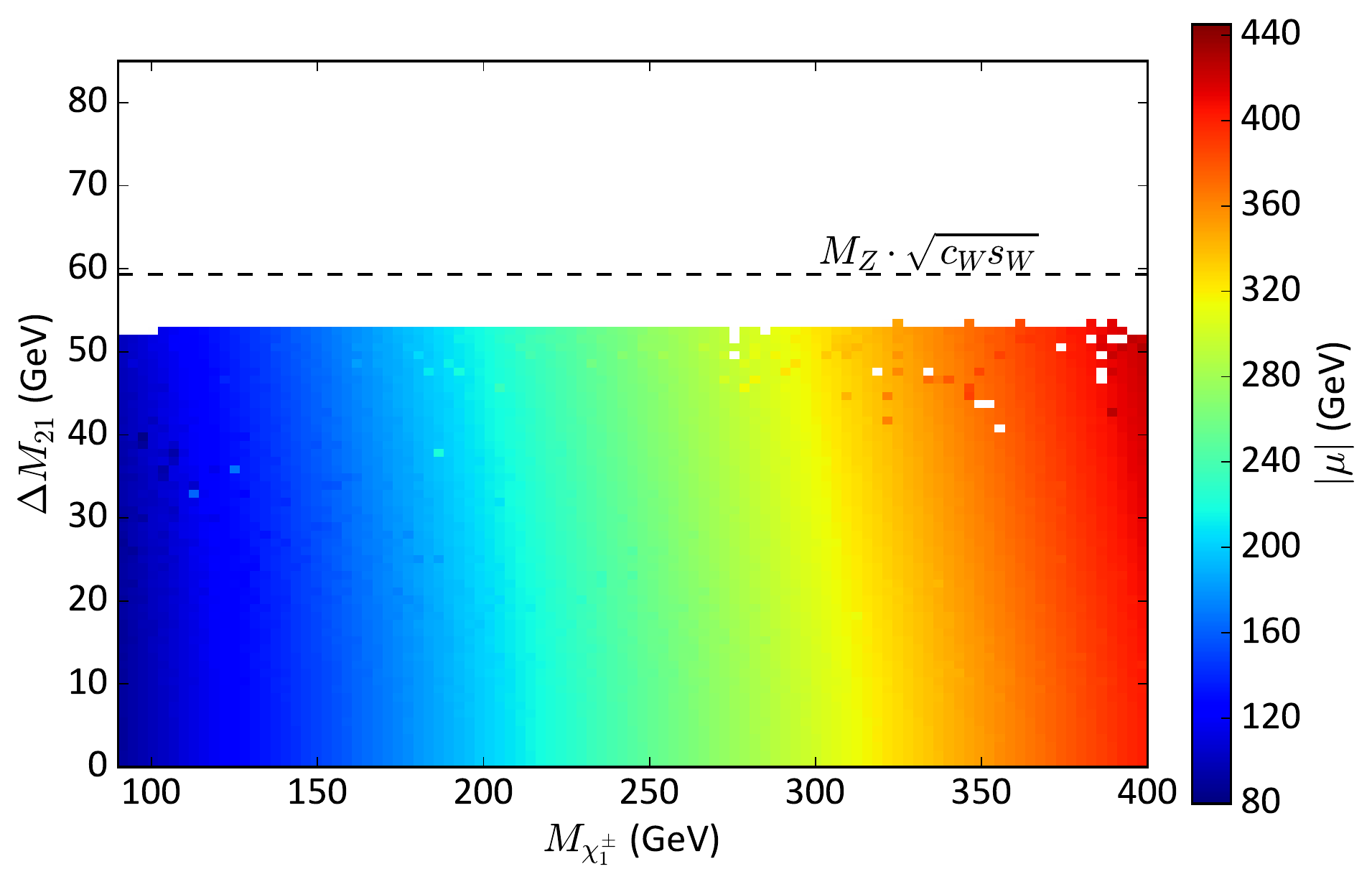}} \\[-4mm]
   \subfigure[\mbox{$\tan \beta$ ($\mu > 0$)}]
  {\label{fig:tbpar-dista} 
   \includegraphics[width=0.45\textwidth]{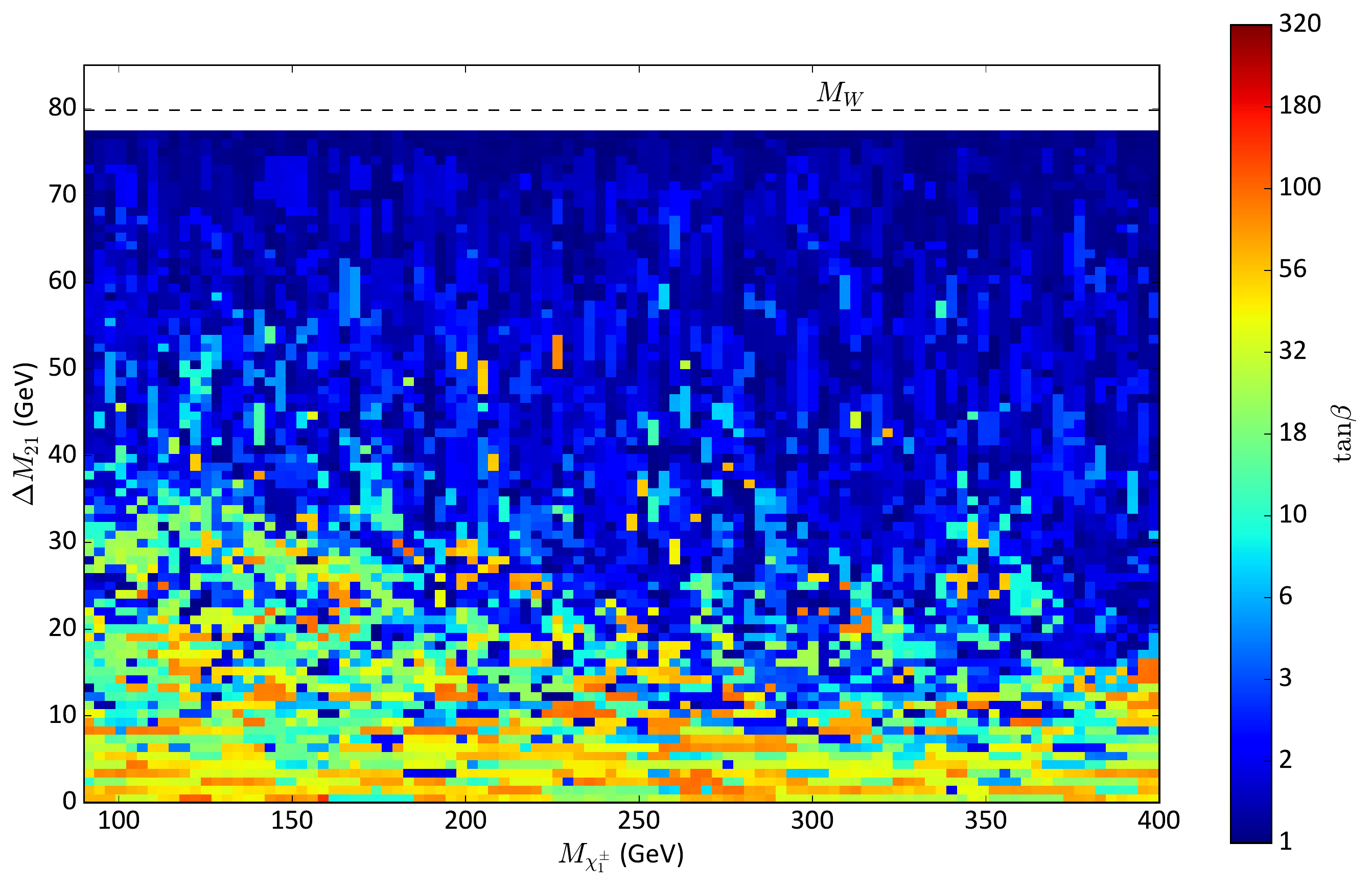}}
   \subfigure[\mbox{$\tan \beta$ ($\mu < 0$)}]
  {\label{fig:tbpar-distb} 
   \includegraphics[width=0.45\textwidth]{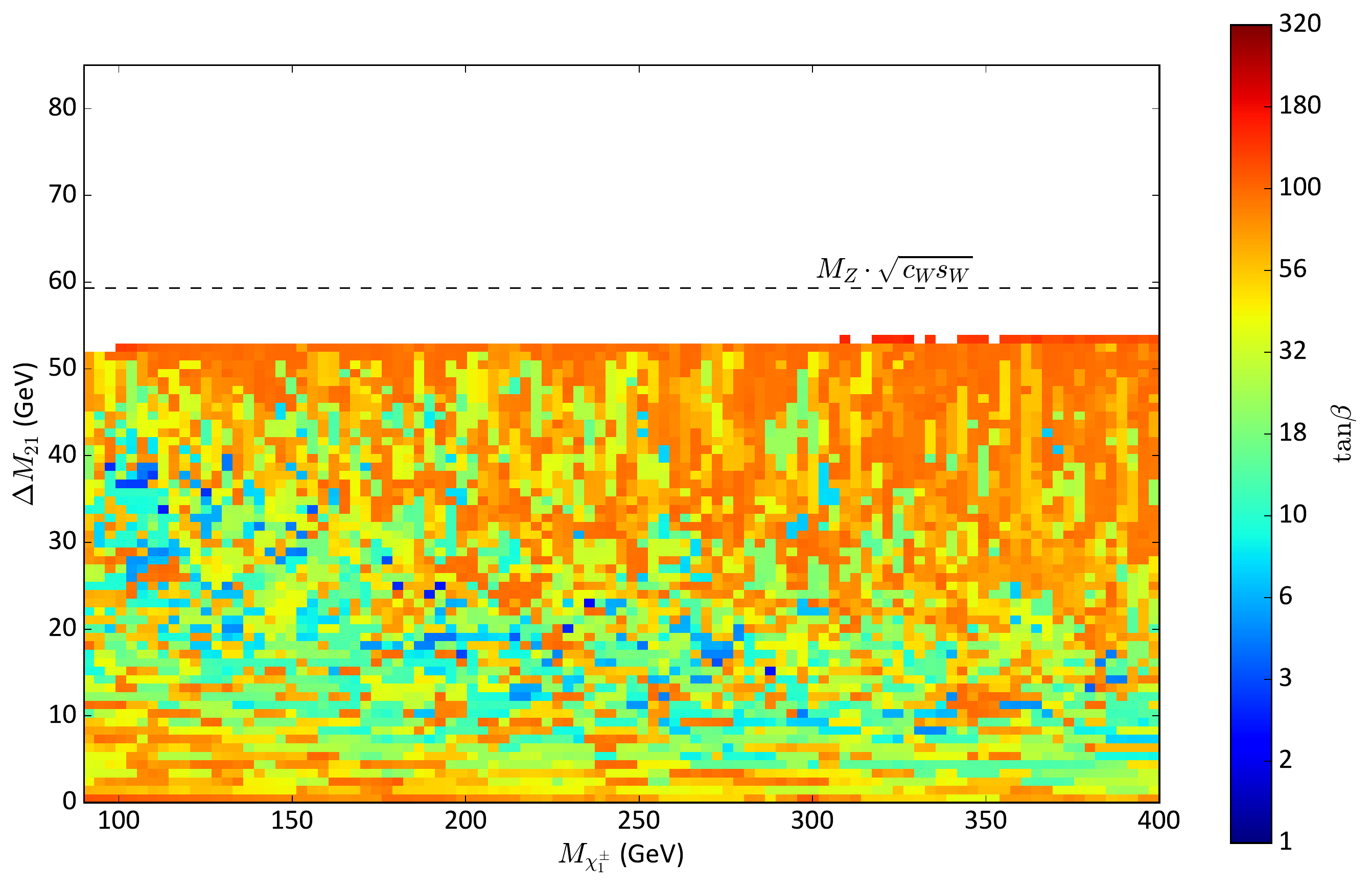}} \\[-4mm]
   \subfigure[\mbox{$M_{1}$ ($\mu > 0$)}]
  {\label{fig:M1par-dista} 
   \includegraphics[width=0.45\textwidth]{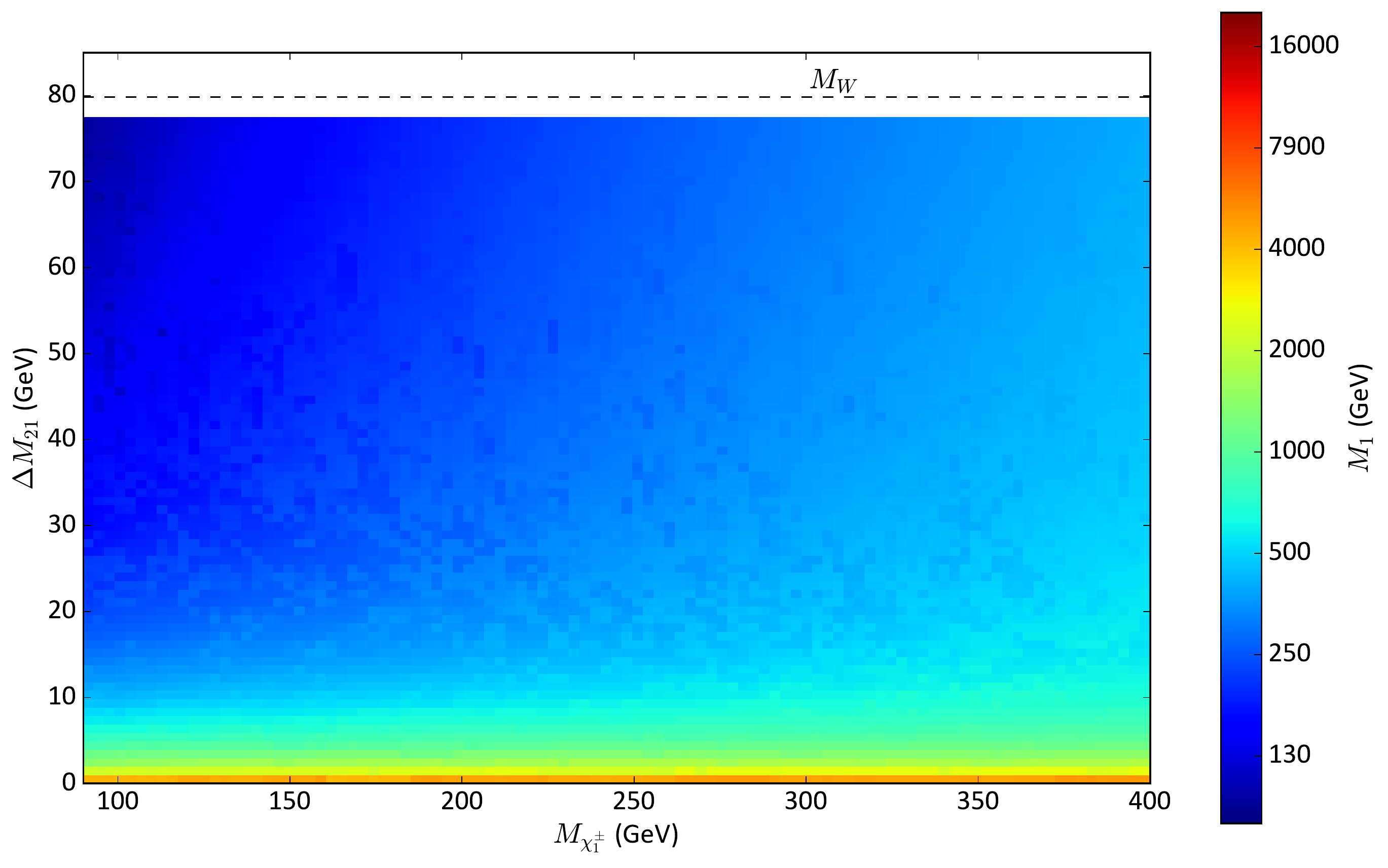}}
   \subfigure[\mbox{$M_{1}$ ($\mu < 0$)}]
  {\label{fig:M1par-distb} 
   \includegraphics[width=0.45\textwidth]{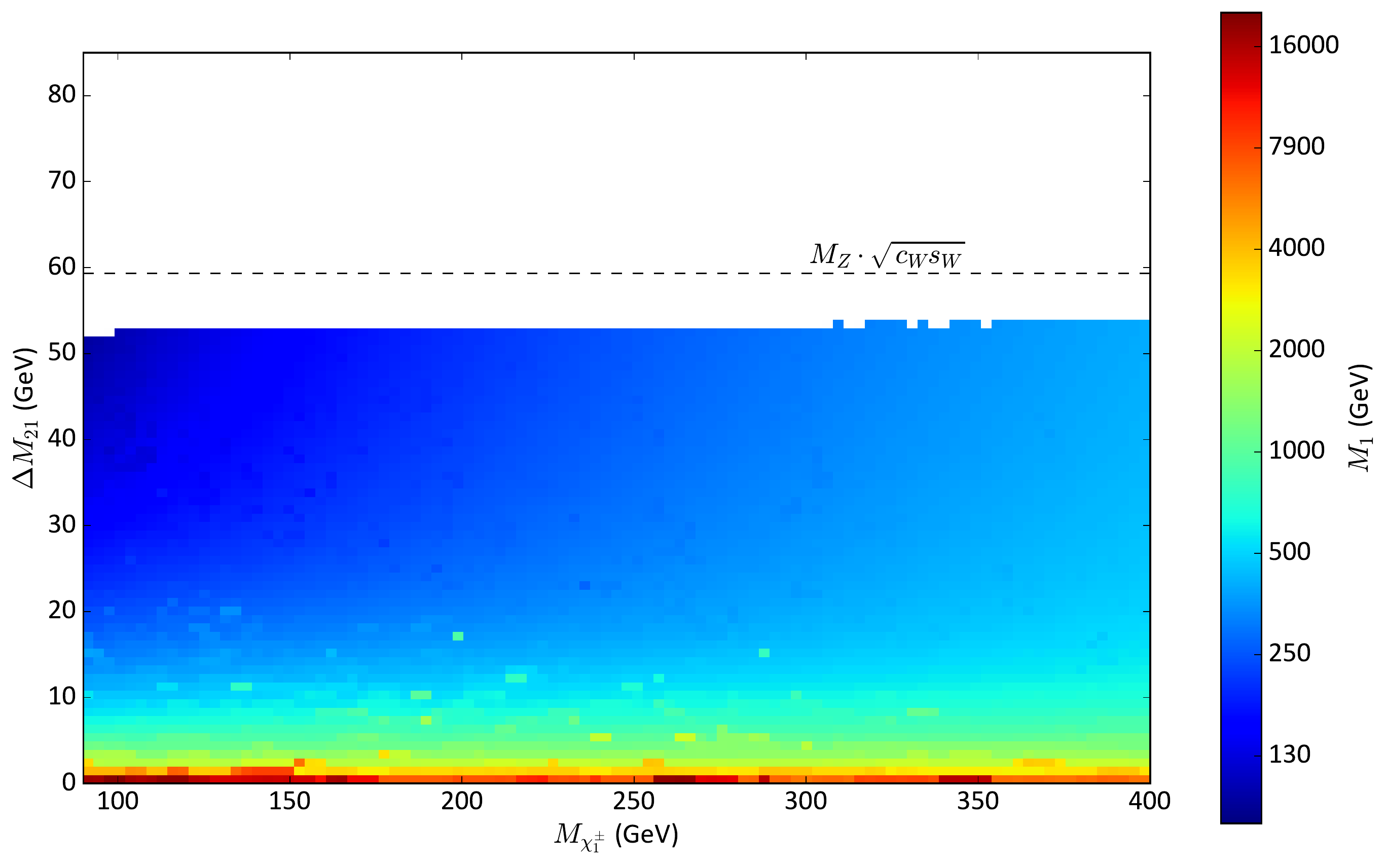}}\\[-4mm]
   \subfigure[\mbox{$M_{2}$ ($\mu > 0$)}]
  {\label{fig:M2par-dista} 
   \includegraphics[width=0.45\textwidth]{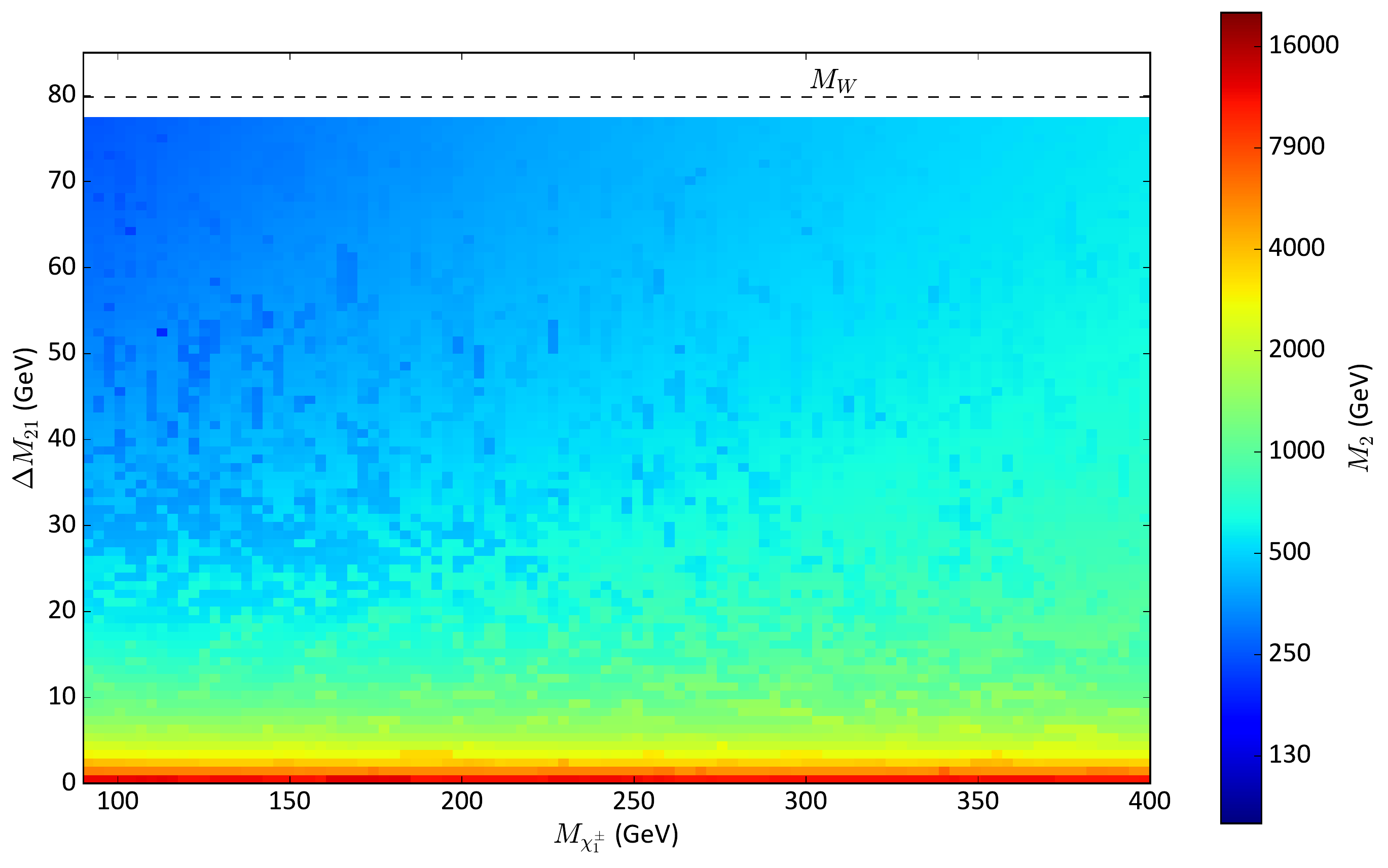}}
   \subfigure[\mbox{$M_{2}$ ($\mu < 0$)}]
  {\label{fig:M2par-distb} 
   \includegraphics[width=0.45\textwidth]{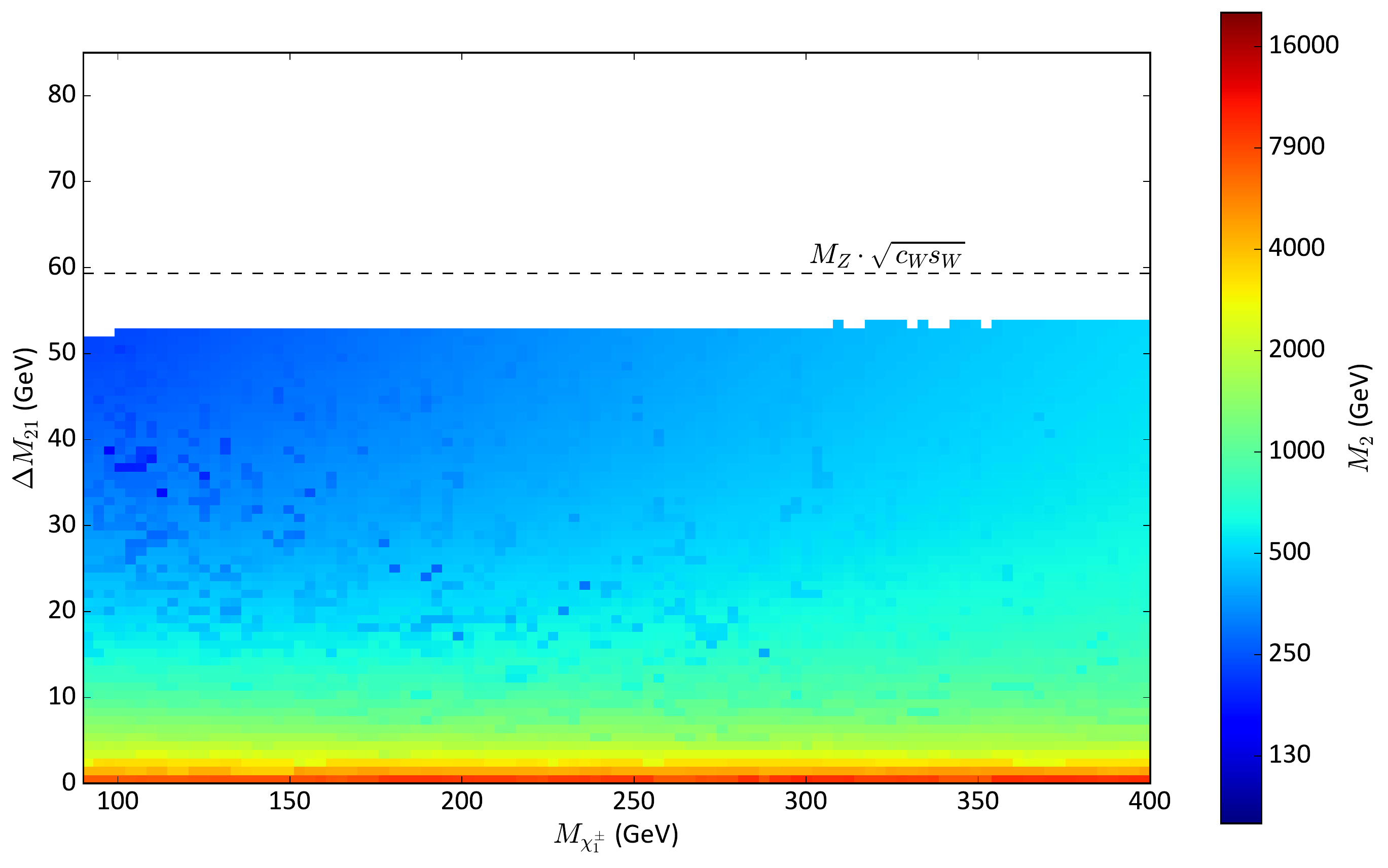}}
    \end{center}
 \caption{\label{fig:5}Fitted MSSM parameters $|\mu|$ (a,b), $\tan \beta$ (c,d),
 $M_{1}$ (e,f) and $M_{2}$ (g,h) for $\mu > 0$ and $\mu < 0$, respectively. Due to the limited
 sensitivity of the fit, the last three are shown logarithmically. Chargino masses below
 92.4 GeV are excluded by LEP in the higgsino region for any lightest neutralino mass.
 This limit rises to 103.5 GeV for mass splittings larger than 5 GeV \cite{lepsusy,%
 particledatagroup:2014}.
}
\end{figure*}

The score distributions in Fig.\ \ref{fig:firsteval} indicate that in our specific case study,
the mass splittings between light higgsinos should not exceed $M_W$ for $\mu > 0$ and $M_Z \cdot
\sqrt{s_{W} c_{W}}$ for $\mu < 0$. Large neutralino mass splittings $\Delta M_{21}$ mostly entail
higgsino contents of less than 70\% and as low as $\sim 50\%$ for the largest values of
$\Delta M_{21}$.
%
This result is nearly independent of the physical chargino mass $M_{\chi^{\pm}_{1}}$.

In Fig.\ \ref{fig:quanteval}, we therefore show the higgsino content as a function of
$\Delta M_{21}$ only for both $\mu>0$ (blue crosses) and $\mu<0$ (red diamonds). We find that it
falls off quadratically for mass splittings below roughly 25 GeV. For positive values of $\mu$,
the fall-off then becomes linear beyond this value.

\subsection{MSSM scenarios} \label{ssec:4.3}

In Fig.\ \ref{fig:5}, we display the fitted MSSM parameters $|\mu|$ (a,b), $\tan
\beta$ (c,d), $M_{1}$ (e,f) and $M_{2}$ (g,h) for $\mu > 0$ and $\mu < 0$, respectively. Due to the
limited sensitivity of the fit, the last three are shown logarithmically. In the following
discussion of these figures, we focus on general trends, exceptional behaviour, and the amount
of fine-tuning that is necessary to reproduce the desired mass spectrum and higgsino content.
Statistically, fine-tuned models are unlikely to be realised in nature and are often taken
as a hint for new, so far poorly understood symmetries in physics.

The $\mu$-parameter distributions found for $\mu < 0$ and $\mu > 0$ are shown in Fig.\
\ref{fig:5} (a,b), respectively. They confirm our hypothesis that $|\mu|$ is mostly fixed by the
chargino mass $M_{\chi^{\pm}_{1}}$. The additional dependence on $\Delta M_{21}$ is characterised
by
\beq
 \label{eq:mu_reex} 
 |\mu | = M_{\chi^{\pm}_{1}} + \frac{\Delta M_{21}}{2}+\epsilon_{\mu} M_{W} \ ,
\eeq
where $\epsilon_{\mu}$ parameterises the deviations from these linear dependencies that are at
most of ${\cal O}(M_W)$. For 95\% of the models, $\epsilon_{\mu}$ lies within $\lb[-0.09,
0.08\rb]$ for $\mu > 0$ and $\lb[-0.17,0.0\rb]$ for $\mu < 0$ with the largest deviations found
in the region $\mu < 0$, $\Delta M_{21} \gtrsim 20 \ \text{GeV}$ and $M_{\chi^{\pm}_{1}} \lesssim 170
\ \text{GeV}$. The two leading terms in Eq.\ (\ref{eq:mu_reex}) are thus accurate about $10-20$
GeV.

The $\tan \beta$-parameter distributions found for $\mu < 0$ and $\mu > 0$ are shown in Fig.\
\ref{fig:5} (c,d), respectively. A general, though weak trend is that one obtains smaller
$\tan\beta$ for larger $\Delta M_{21}$ for $\mu > 0$, but larger $\tan\beta$ for larger $\Delta
M_{21}$ for $\mu<0$. This corresponds to the opposite dependencies on $\tan\beta$ observed in
the upper right parts of Figs.\ \ref{fig:mupos} and \ref{fig:muneg} in Sec.\ \ref{sec:2}.
The weak dependence of the spectrum for large values of $\tan\beta$ has been discussed before.
As we can observe now, it appears in particular for $\mu<0$ or $\mu>0$ and small $\Delta M_{21}$,
while the allowed range of $\tan\beta$ becomes more limited for $\mu >0$ and large neutralino
mass splittings.

In Fig.\ \ref{fig:5} (e-h), the distributions of the gaugino mass parameters $M_{1}$ and $M_{2}$
are shown for both $\mu > 0$ and $\mu < 0$. The distributions of both parameters vary by almost
two orders of magnitude and roughly inversely to the neutralino mass splitting $\Delta M_{12}$.
We parameterise the fitted gaugino mass parameters $M_1$ and $M_2$ by
\beq
 \label{eq:mi_reex} 
M_{1,2} = M_{\chi^{\pm}_{1}} - \frac{\Delta M_{21}}{2} + \epsilon_{M_{1,2}}\frac{M_{W}^{2}}{\Delta M_{21}} \ .
\eeq
This expression does not reproduce the correlation of $\Delta M_{21}$ and $\tan\beta$ discussed
above and is thus less accurate than our parameterisation of $|\mu|$ in Eq.\ (\ref{eq:mu_reex}).
In particular, the parameters $\epsilon_{M_{1,2}}$, that parameterise deviations from the two
leading terms, lie in the large ranges $\epsilon_{M_{1,2}}\in[0.18,0.82],[1.26,2.42]$ for
$\mu>0$ and $\epsilon_{M_{1,2}}\in[0.10,3.05],[0.82,1.43]$ for $\mu < 0$ and for 95\% of the
models. This is due to the known fact that pure higgsinos, often
associated with $M_{1,2}\gg\mu$,
have small mass splittings, so that the requested spectrum is not very sensitive to the exact
values of the gaugino masses.
The region of $|\mu|\simeq M_1$, but $M_2\gg|\mu|$, known as the ``well-tempered
bino/higgsino'' region to identify the resulting light gaugino states as mixed binos and
higgsinos, allows for the approximate analytic diagonalization of the reduced
three-dimensional neutralino mass matrix \cite{ArkaniHamed:2006mb}. We already
observed this increased mixing in Fig.\ \ref{fig:quanteval}, where the higgsino
content fell linearly for sizeable mass splittings $\Delta M_{21}$.

\begin{figure*}
   \begin{center} 
   \subfigure[\mbox{fine-tuning ($\mu > 0$)} ]{\label{fig:finetune-1a} 
   \includegraphics[width=0.485\textwidth]{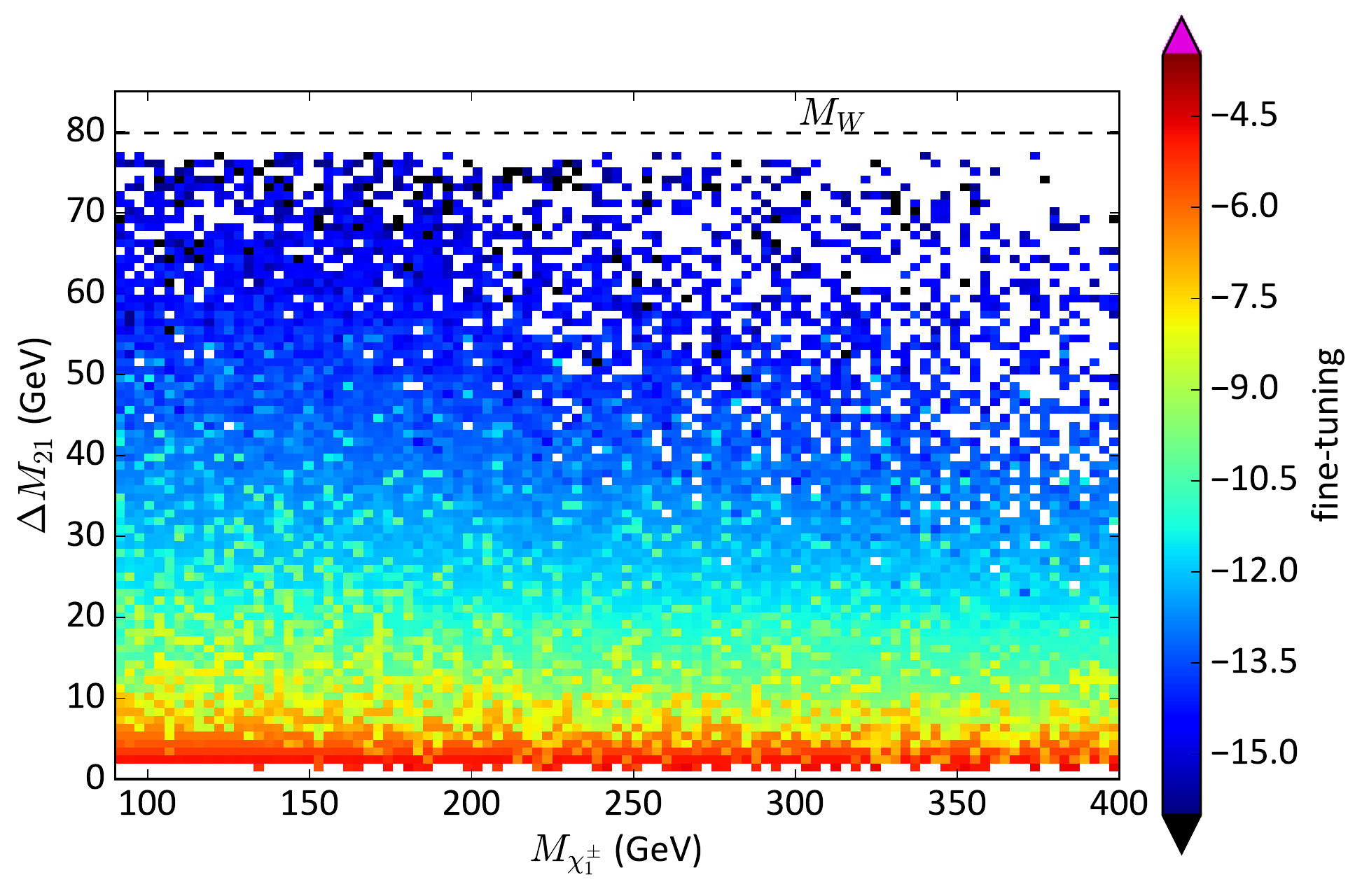}}
   \subfigure[\mbox{fine-tuning ($\mu < 0$)} ]{\label{fig:finetune-1b} 
   \includegraphics[width=0.49\textwidth]{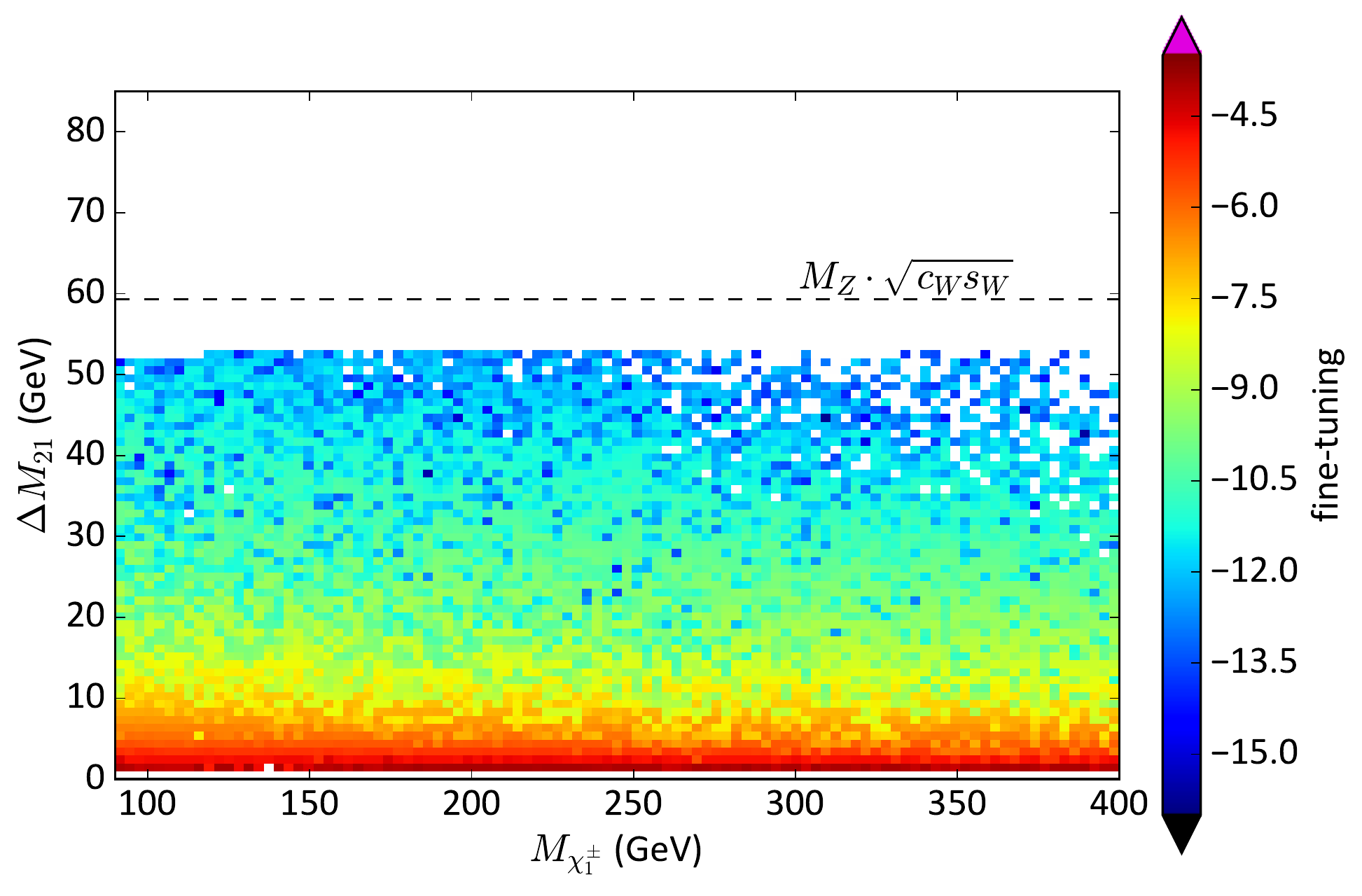}}\\[-3mm]
   \subfigure[\mbox{number of boundary violations ($\mu > 0$)} ]{\label{fig:finetune-2a} 
   \includegraphics[width=0.485\textwidth]{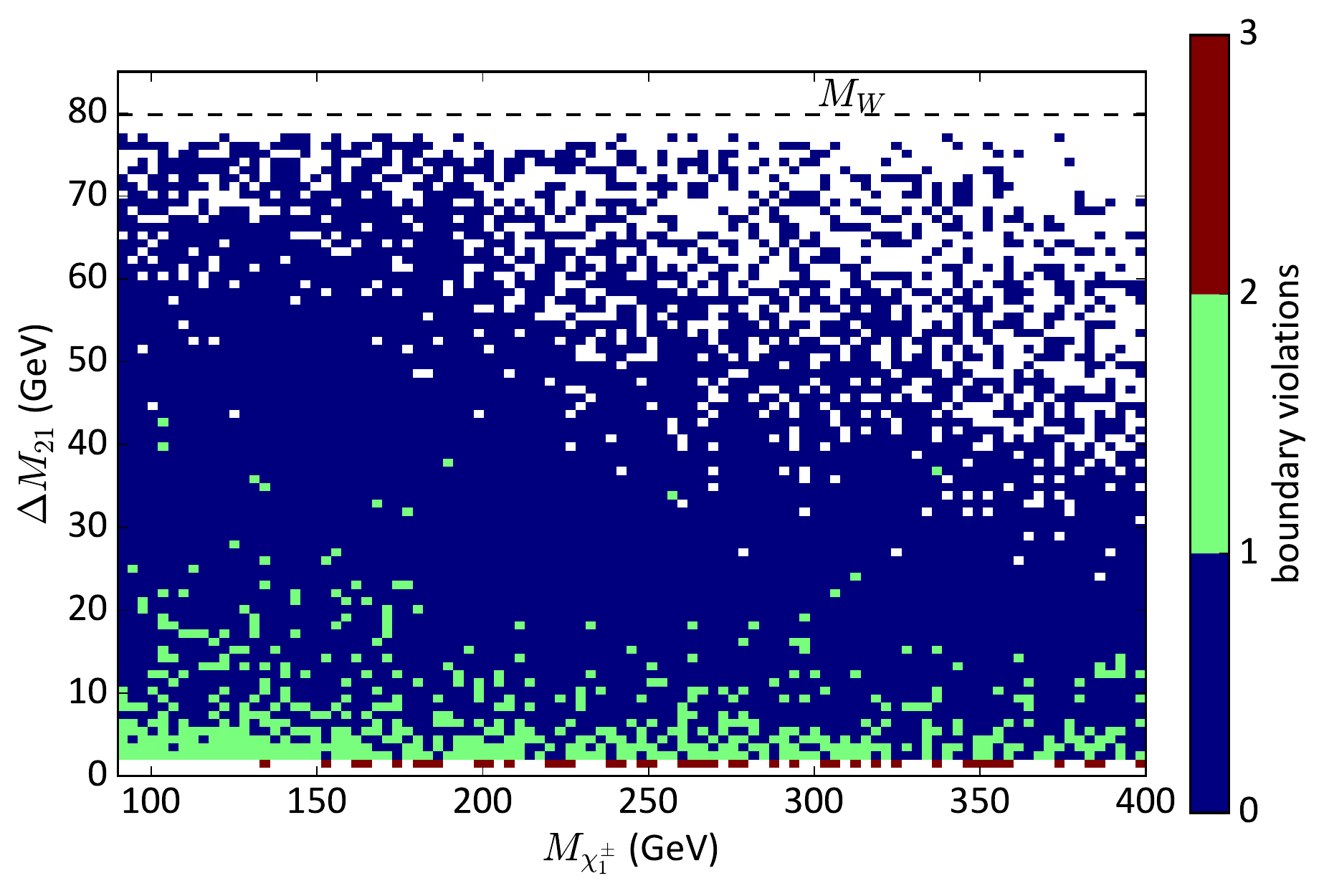}}
   \subfigure[\mbox{number of boundary violations ($\mu < 0$)}  ]{\label{fig:finetune-2b} 
   \includegraphics[width=0.49\textwidth]{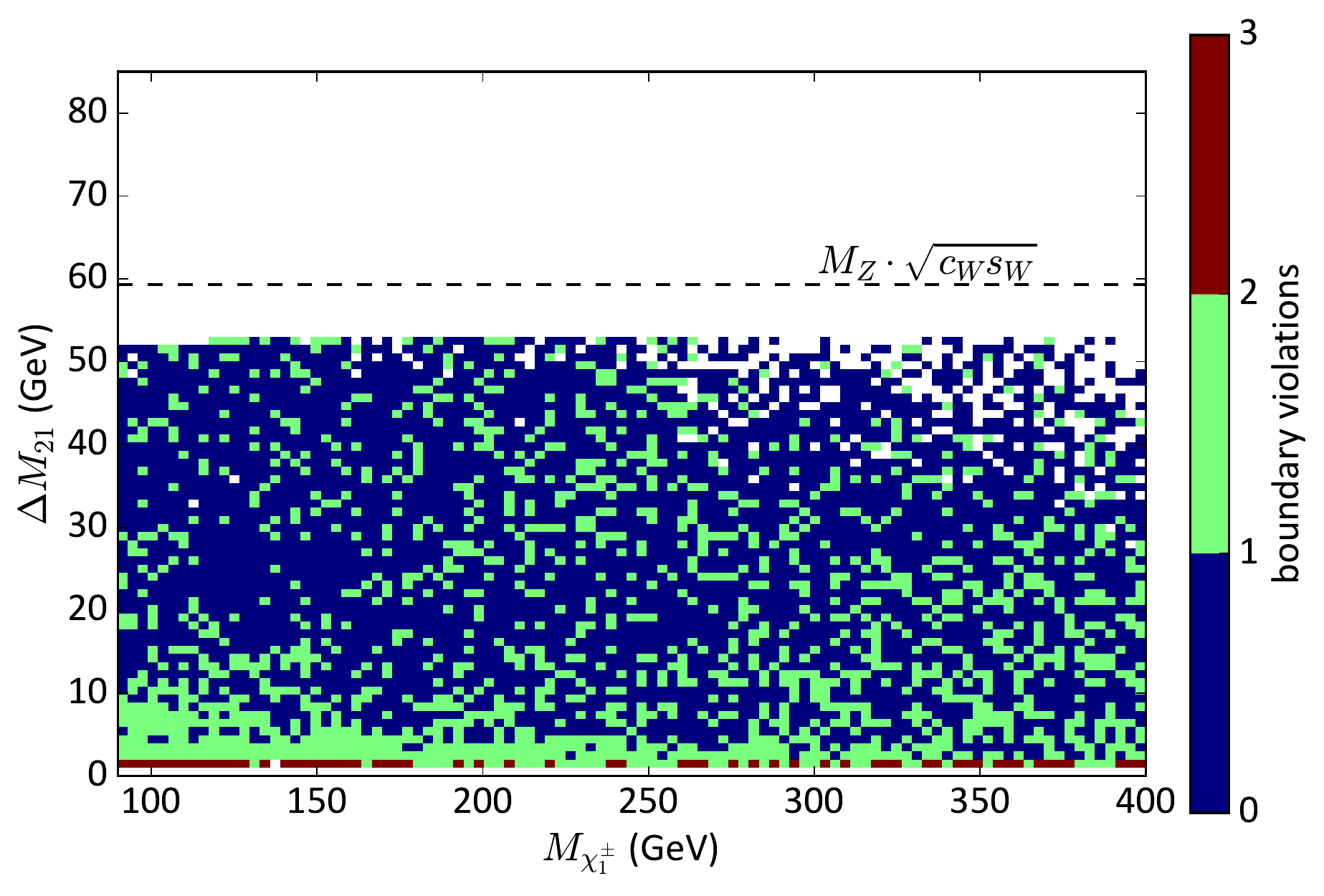}}
    \end{center}
 \caption{\label{fig:finetune}Upper figures: Logarithmic representation of the fine-tuning level of
 simplified light higgsino models in terms of the relative acceptable ranges of the underlying
 MSSM parameters for $\mu>0$ (a) and $\mu<0$ (b). Large negative numbers therefore indicate
 large fine-tuning. Lower figures: Number of boundary violations of the initial parameter
 ranges for $\mu>0$ (c) and $\mu<0$ (d). Chargino masses below
 92.4 GeV are excluded by LEP in the higgsino region for any lightest neutralino mass.
 This limit rises to 103.5 GeV for mass splittings larger than 5 GeV \cite{lepsusy,%
 particledatagroup:2014}.}
\end{figure*}

Models that reproduce similar physical masses or mixings, but originate from
very different, sometimes isolated fundamental parameters, signal the presence of fine-tuning.
We quantify this fine-tuning by multiplying for each benchmark the variations of the
fundamental parameters $|\mu|$, $M_1$, $M_2$ and $\tan\beta$ leading to acceptable scores
(below 0.1) and
then dividing by the corresponding total ranges as defined in Eq.\ (\ref{eq:scanranges-1}).
The result is shown in Fig. \ref{fig:finetune} (a,b) for positive and negative values of $\mu$,
respectively. Due to the logarithmic representation, large negative numbers correspond to large
fine-tuning. It occurs more often for large mass splittings and/or positive values of $\mu$
confirming that conversely pure higgsino scenarios usually have small mass splittings and
are then less sensitive to specific choices e.g. of $M_1$, $M_2$ or $\tan\beta$.
Furthermore, in some cases acceptable models also lie outside the parameter ranges given in
Eq.\ (\ref{eq:scanranges-1}). The number of such initial boundary violations is displayed in
Fig.\ \ref{fig:finetune} (c,d), again for $\mu>0$ and $\mu<0$, respectively. As one can see,
acceptable models in larger regions of the parameter space exist often for (nearly) mass
degenerate light neutralinos, where large $M_{1,2}$ allow for compressed higgsino mass spectra,
and in the case $\mu<0$, where $\tan\beta$ can be very large.

\subsection{The Higgs-stop sector} \label{ssec:4.5}

The large MSSM parameter space allows one (at least at tree-level) to decouple squarks, gluinos
and sleptons without any impact on the gaugino-higgsino sector. Care is, however, required for the
decoupling of the higgs-stop sector due to the large impact of stop radiative corrections on the
mass of the observed SM-like Higgs boson, that has to match the measured value of 125 GeV.
In the absence of stop mixing, the squared $CP$-even and $CP$-odd neutral Higgs boson masses
are related to the top quark mass $m_t$ and the stop mass $m_{\tilde t}$ through \cite{Drees:873465}
\beq
 \label{eq:higgscorr-stop} 
 m^{2}_{h^{0}} + m^{2}_{H^{0}} = m^{2}_{A^{0}}+M^{2}_{Z}+\frac{3G_{F}m^{4}_{t}}{\sqrt{2}\pi^{2} s^{2}_{\beta}} \ln \frac{m^{2}_{\tilde{t}}}{m^{2}_{t}},
\eeq
where $G_F$ is the Fermi constant. This entails
\beq
 m^{2}_{\tilde{t}} \simeq m^{2}_{t} \exp\lb(\frac{\sqrt{2}\pi^{2} s^{2}_{\beta}}{3G_{F}m^{4}_{t}}\lb(m^{2}_{h^{0}}-M^{2}_{Z}\rb)\rb)
\eeq
for $m_{H^0}\simeq m_{A^0}$ or $m_{\tilde{t}} \in [ 885 \text{ GeV}, 1330 \text{ GeV}]$, as long as
$\tan\beta>2$. The full additional parameter space for the Higgs-stop sector includes the
squared off-diagonal Higgs mass parameter $m_{12}^2$, the soft SUSY-breaking mass parameters
$m^{2}_{\tilde{Q}_{3}}$ and $m^{2}_{{\tilde t}_R}$, and the trilinear coupling $A_{t}$, all taken to be
real to avoid new sources of $CP$-violation. A scan over this additional parameter space for a
given MSSM higgsino model with fixed $\mu$ and $\tan\beta$ and full stop mixing leads to a
successful decoupling of the heavy Higgs bosons. The corresponding
regions in the $CP$-odd neutral Higgs mass and the physical stop masses are
\be\bsp
 m_{A^0}\in[992,4386]~{\rm GeV}~(\mu>0)\ , \\
 m_{{\tilde t}_1}\in[752,1481]~{\rm GeV}~(\mu>0)\ ,\\
 m_{{\tilde t}_2}\in[1607,2487]~{\rm GeV}~(\mu>0)\ ,\\
\esp\ee
and
\be\bsp
 m_{A^0}\in[1063,3925]~{\rm GeV}~(\mu<0)\ ,\\
 m_{{\tilde t}_1}\in[809,1212]~{\rm GeV}~(\mu<0)\ ,\\
 m_{{\tilde t}_2}\in[1840,2413]~{\rm GeV}~(\mu<0)\ .\\
\esp\ee
for positive and negative values of $\mu$, respectively.
As is well known, a light SM-like Higgs boson of mass 125 GeV requires in
general a light stop with a mass below or around 1 TeV and a large stop mass splitting of at
least 1 TeV.

\subsection{Implications on dark matter} \label{ssec:4.5}

\begin{figure*}
   \begin{center} 
   \subfigure[$\log_{10}(\Omega_{\chi}^{\rm MO}/\Omega_{\chi}^{\rm Pl})$ ($\mu > 0$)]
 {\label{fig:cosmo_relicpos}\includegraphics[width=0.49\textwidth]{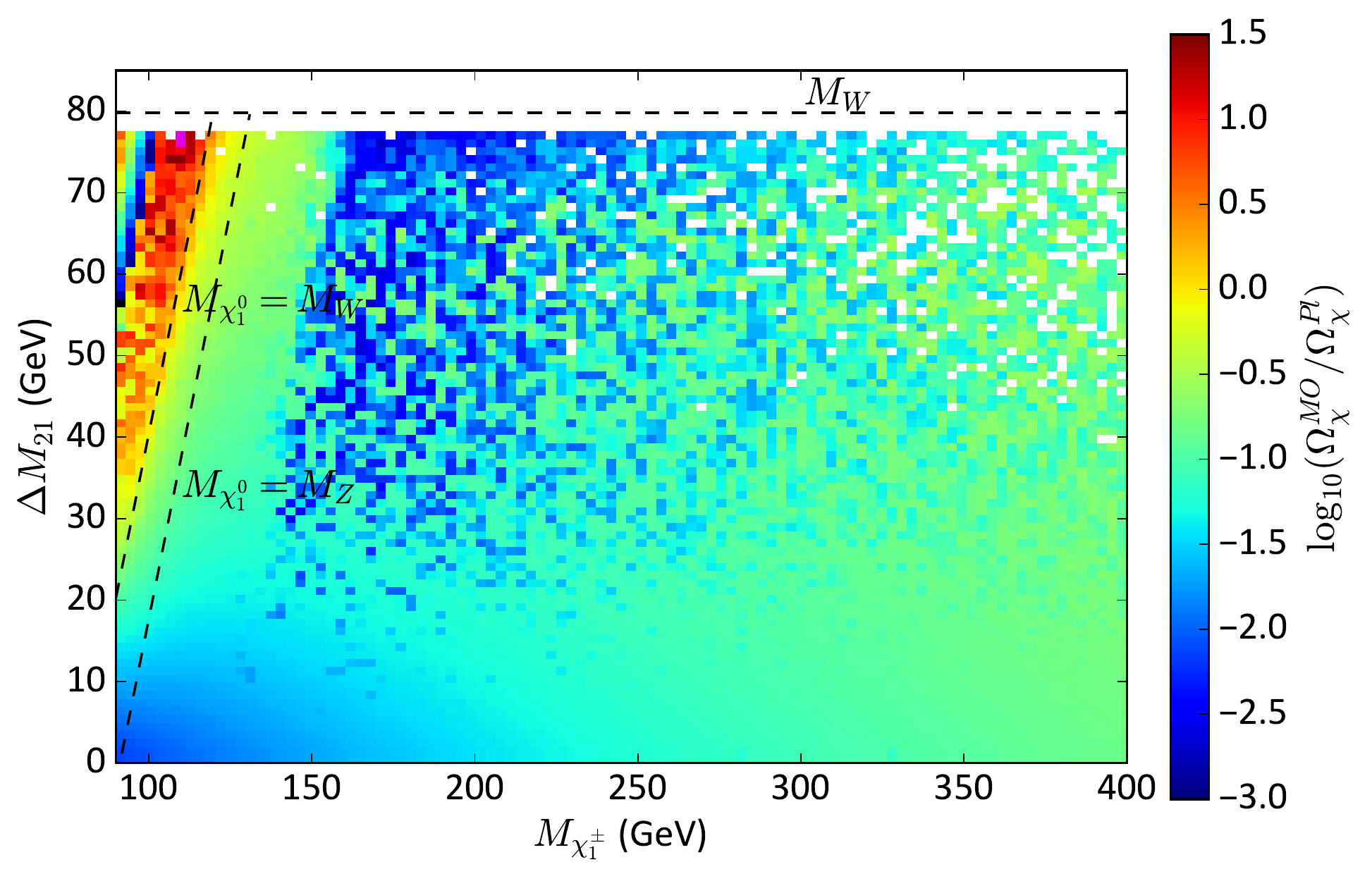}}
   \subfigure[$\log_{10}(\Omega_{\chi}^{\rm MO}/\Omega_{\chi}^{\rm Pl})$ ($\mu < 0$)]
 {\label{fig:cosmo_relicneg}\includegraphics[width=0.49\textwidth]{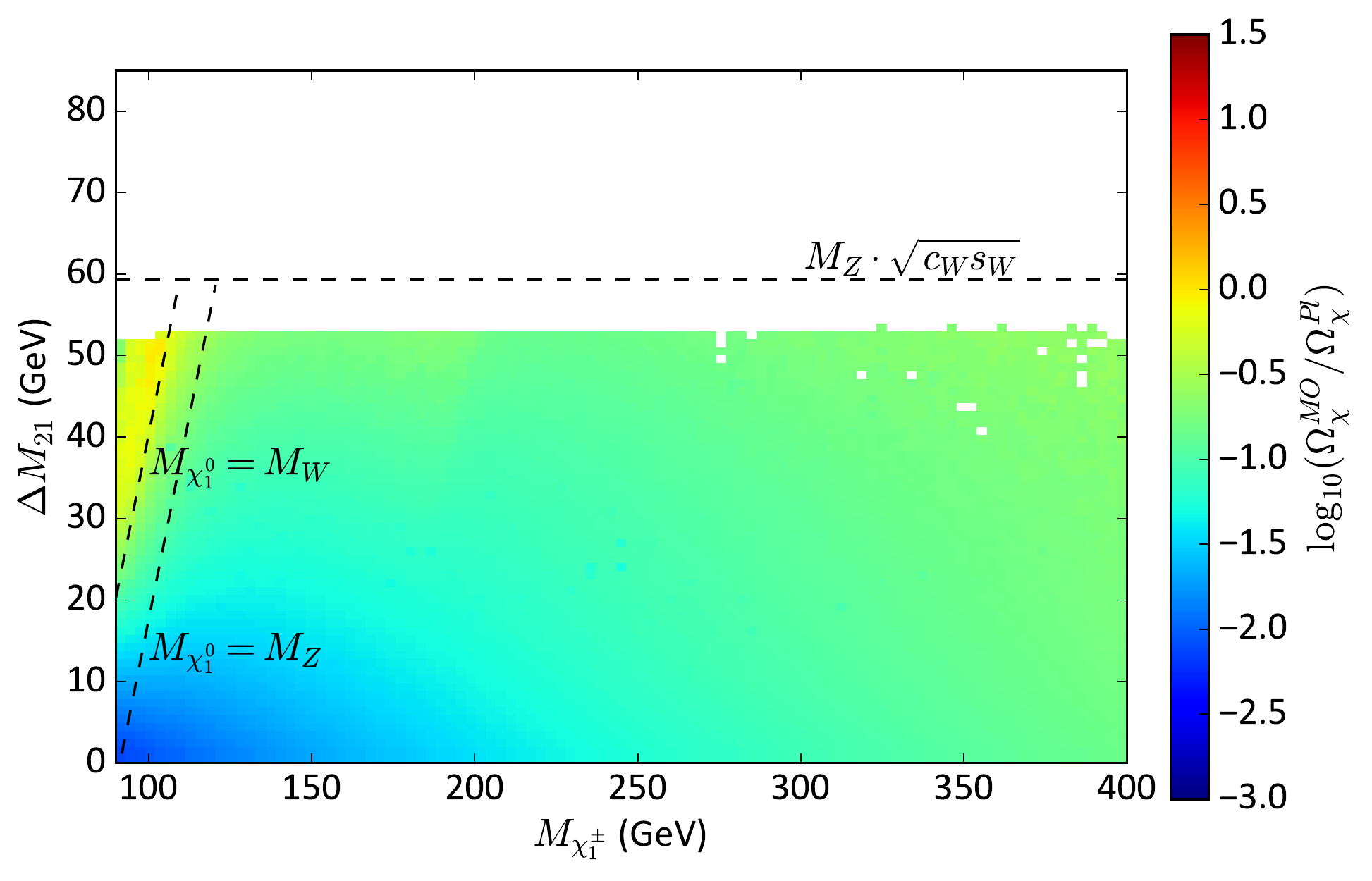}}
 \\[-3mm]
   \subfigure[$\log_{10} (\sigma_{\chi}^{\rm MO})$ ($\mu  > 0$)]
 {\label{fig:cosmo_ddpos}\includegraphics[width=0.49\textwidth]{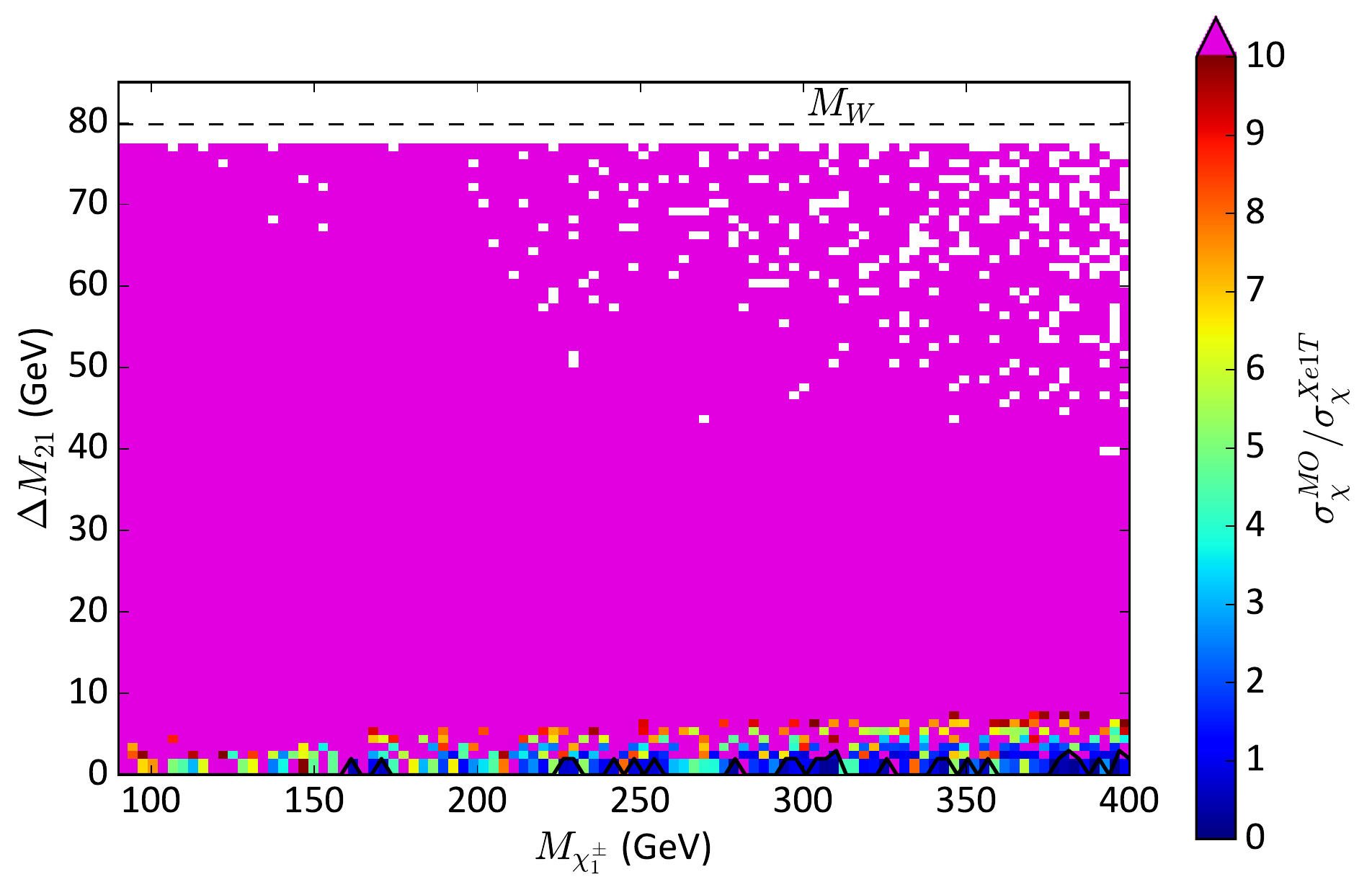}}
   \subfigure[$\log_{10} (\sigma_{\chi}^{\rm MO})$ ($\mu  < 0$)]
 {\label{fig:cosmo_ddneg}\includegraphics[width=0.49\textwidth]{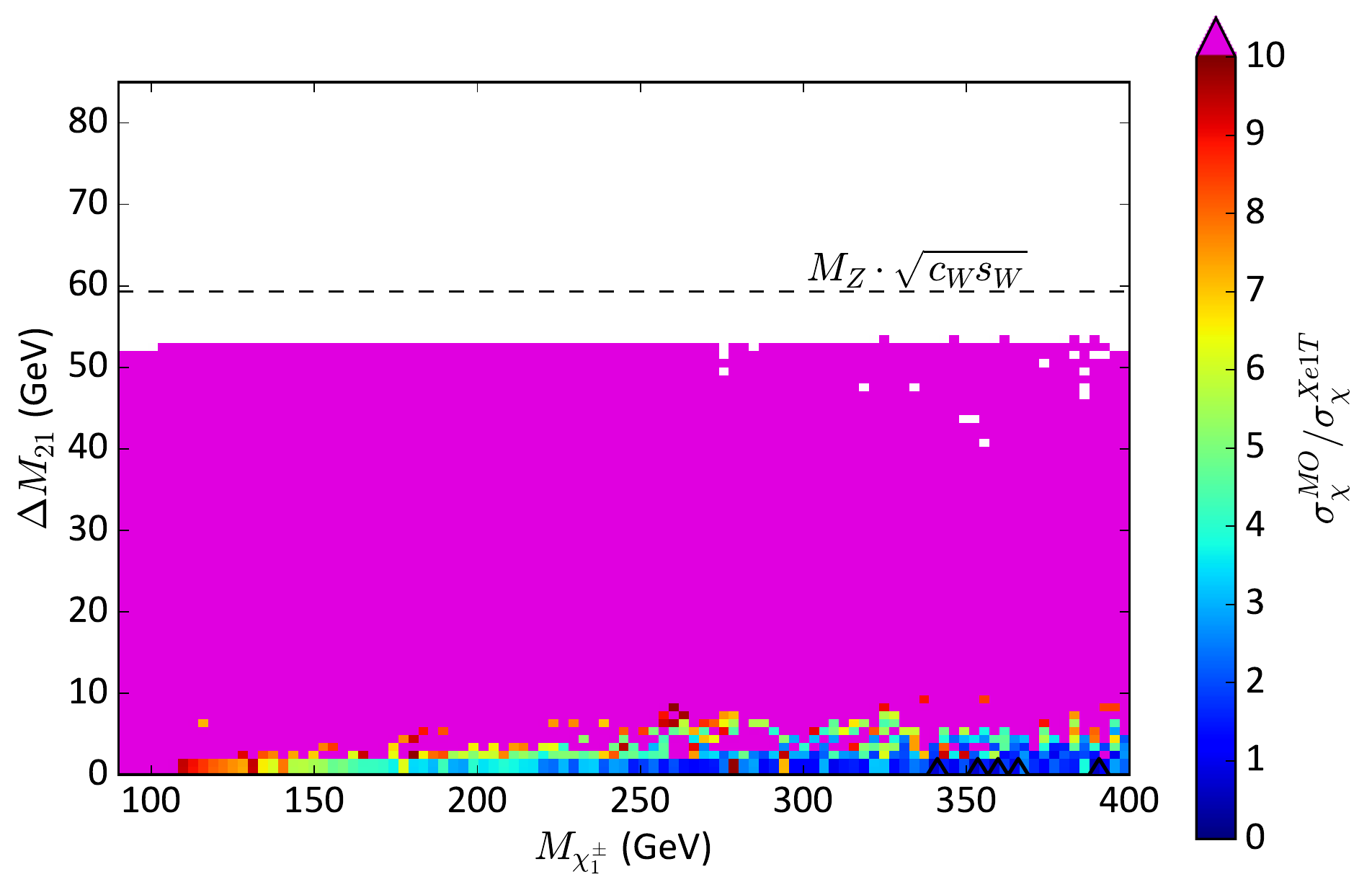}}
    \end{center}
 \caption{\label{fig:cosmo_constraints}Upper figures: Logarithmic ratio of predicted and observed
 relic abundance of light higgsino dark matter for $\mu>0$ (a) and $\mu<0$ (b).
 Lower figures: Ratios of predicted cross sections and Xenon1T exclusion limits
 for the direct detection of higgsino dark matter for $\mu>0$ (c) and $\mu<0$ (d).}
\end{figure*} 

An important motivation for supersymmetry is its prediction of a classic WIMP (weakly interacting
massive particle) dark matter candidate, the lightest neutralino. The relic abundance of dark
matter in the universe has been determined very precisely by the Planck collaboration to be
$\Omega_{\chi}^{\rm Pl}h^{2} = 0.1199 \pm 0.0027$ \cite{Plack2013}. We therefore compare the
relic density $\Omega_{\chi}^{\rm MO}$ predicted for light higgsino MSSM models by the public code
micrOMEGAs \cite{Barducci:2016pcb} to the observed one in Fig.\ \ref{fig:cosmo_constraints}
(a,b) for $\mu>0$ and $\mu<0$. The main observation here is the appearance of the $\chi_{1}^{0}
\chi^{0}_{1}\rightarrow W^{+}W^{-}$ threshold. When this process is kinematically allowed, the
annihilation cross section of $\chi_{1}^{0}$ increases, and therefore the dark matter relic
abundance decreases. Close to (above) this threshold, the cross section is sufficiently small to
explain (at least partially) the measured dark matter relic abundance.
In Fig.\ \ref{fig:cosmo_constraints} (c,d) we show the ratios of predicted direct detection
cross sections over the Xenon1T exclusion limits \cite{Aprile:2017iyp}. In the higgsino mass
range of $M_Z$ to 400 GeV studied here, the cross sections predicted by micrOMEGAs decrease with
the mass splitting from $10^{-6}$ pb to $10^{-11}$ pb. Since the Xenon1T experiment has recently
reached a sensitivity of $\sigma_\chi^{\rm Xe1T}\sim10^{-10}$ pb for WIMP masses of about 100 GeV,
only cross sections below the black line and very small mass splittings are still allowed.
However, since searches at the LHC and in direct detection experiments depend on
different sets of assumptions, both are complementary, and they should both be
taken into account. In the dark matter context, the potential
LHC constraints on the considered models include both results from direct
searches for supersymmetry and for dark matter in general, like when using
monojet probes that are expected to be golden handles on compressed
electroweakino spectra~\cite{Schwaller:2013baa}.
Models with light gravitinos imply, of course, a very
different dark matter phenomenology.

\section{Conclusion} \label{sec:5}

Simplified SUSY models have become a popular tool for model-independent searches at the LHC.
Recently, the LHC experiments ATLAS and CMS have also applied this approach to light neutralinos
and charginos with predefined physical mass spectra and pure gaugino or higgsino content.
We have emphasised in this paper that these models can violate physical principles such as
supersymmetry, gauge invariance, or the consistent combination of production cross sections
and decay branching ratios and that they must therefore be embedded in full MSSM models, whose
relevant four-dimensional parameter space is spanned by $\mu$, $\tan\beta$, $M_1$ and $M_2$.

Exploiting the symmetries of the neutralino and chargino mass matrices, we diagonalised them
and discussed the leading and sub-leading dependencies of the resulting physical mass spectra
and decompositions on these parameters. We then devised an efficient scan strategy for the
full parameter space given a desired physical mass spectrum and introduced a measure for the
quality of our full MSSM reproduction of this spectrum, that could also include criteria such
as a maximal gaugino or higgsino component or couplings to specific sparticles. As a case
study, we investigated the MSSM realisations of light higgsinos, finding an upper bound on the
possible mass splitting among the lightest neutralinos of ${\cal O}(M_W)$ and a lower bound on
the higgsino
content of about 70\%. We saw that large mass splittings required a more substantial level
of fine-tuning, whereas for small mass splittings even larger regions of parameter space than
those scanned by us led to viable scenarios. As expected, squarks, gluinos, and sleptons could
be decoupled to 1.5 TeV, as could the heavier Higgs bosons without spoiling the reproduction
of a SM-like light Higgs boson of mass 125 GeV. The latter required, however, a light stop
of mass below or around 1 TeV with its heavier partner split by at least 1 TeV. The observed
dark matter relic density could be reproduced close to the threshold of neutralino annihilation
into pairs of $W$-bosons, whereas for higher masses the higgsinos can only represent a fraction
of the observed dark matter. The corresponding direct detection cross sections are within reach
of current experiments such as Xenon1T.

While we have indicated how our strategy can be generalised to other scenarios such as those
with non-equidistant mass splitting of the light neutralinos and chargino or those with specific
couplings of gauginos, higgsinos and other sparticles, specific studies of these other scenarios
are beyond the scope of the present work and should be performed with a detailed application in
mind.

\section*{Acknowledgements}
We thank W.\ Adam, C.\ Heidegger, B.\ Schneider and L.\ Shchutska for useful discussions. This
work has been supported by the ANR under contracts ANR-11-IDEX-0004-02 and ANR-10-LABX-63, the
BMBF under contract 05H15PMCCA, the CNRS under contract PICS 150423, and the DFG through the
Research Training Network 2149 ``Strong and weak interactions - from hadrons to dark matter''.

\bibliographystyle{JHEP}
\bibliography{biblio}
\end{document}